\documentclass[%
 reprint, 
 amsmath,amssymb,
 aps, prd, longbibliography, nofootinbib, floatfix
]{revtex4-1}

\usepackage{silence}
\usepackage{graphicx}
\usepackage{dcolumn}
\usepackage{bm}
\usepackage{tabularx}
\usepackage{natbib}
\usepackage{lineno}
\usepackage [english]{babel}
\usepackage [autostyle, english = american]{csquotes}
\MakeOuterQuote{"}
\usepackage{hyperref}
\usepackage{relsize}
\usepackage{float}
\usepackage{xcolor}

\def\DeltanuRF{\Delta\nu_{\lower1.5pt\hbox{$\scriptstyle \mathrm{RF}$}}}
\def\Eobs{\mathbf{E}_{\lower0.5pt\hbox{\scriptsize obs}}}
\def\kB{k_ {\lower1.5pt\hbox{$\scriptstyle B$}}}

\def\rhoDM{\rho_{\lower1.5pt\hbox{\scriptsize DM}}}
\def\sigmaP{\sigma_{\kern -1pt                   
                  \lower1.5pt\hbox{$\scriptstyle P$}}}
\def\sigmaT{\sigma_{\lower1.5pt\hbox{$\scriptstyle T$}}}

\begin{document}


\title{New Limit on Dark Photon Kinetic Mixing in the 0.2-1.2 \texorpdfstring{$\boldsymbol{\mu}$}{u}eV Mass Range \break From the Dark E-Field Radio Experiment}

\author{Joseph Levine}

\author{Benjamin Godfrey}

\author{J. Anthony Tyson}
 \email{tyson@physics.ucdavis.edu}


\author{S. Mani Tripathi}

\author{Daniel Polin}

\author{Amin Aminaei}

\affiliation{Physics and Astronomy Department, UC Davis, Davis, California, 95616 USA}

\author{Brian H. Kolner}

\affiliation{Electrical and Computer Engineering Department, UC Davis, Davis, California, 95616, USA \\
Physics and Astronomy Department, UC Davis, Davis, California, 95616 USA}

\author{Paul Stucky}
\affiliation{Chemistry Department, UC Davis, 
Davis, California, 95616, USA}%

\date{\today}

\begin{abstract}
We report new limits on the kinetic mixing strength of the dark photon spanning the mass range 0.21 -- 1.24\,$\mu$eV corresponding to a frequency span of 50 -- 300\,MHz. The Dark E-Field Radio experiment is a wide-band search for dark photon dark matter. 
In this paper we detail changes in calibration and upgrades since our proof-of-concept pilot run. Our detector employs a wide bandwidth E-field antenna moved to multiple positions in a shielded room, a low noise amplifier, wideband ADC, followed by a $2^{24}$-point FFT. An optimal filter searches for signals with Q~$\approx10^6$. In nine days of integration, this system is capable of detecting dark photon signals corresponding to $\epsilon$ several orders of magnitude lower than previous limits. We find a 95\% exclusion limit on $\epsilon$ over this mass range between $6\times 10^{-15}$ and $6\times 10^{-13}$, tracking the complex resonant mode structure in the shielded room.
\bigskip

\end{abstract}
\maketitle



\section{Introduction}
\label{sec:intro}

Dark matter was discovered via its gravitational effects on large scale dynamics of galaxies and stars within galaxies. Indeed, galaxies are held together by their dark matter halos, which supply much more mass than the baryons (stars and gas). While abundant astrophysical and astronomical observations have characterized the gravitational interactions of dark matter over a variety of length and time scales, no other interactions have been conclusively detected. Recently, searches have largely focused on the WIMP hypothesis and found nothing \cite{misiaszek2024direct}. However, the parameter space of dark matter is vast, motivating the need for alternative theories and detection techniques. The Snowmass 2021 Cosmic Frontier report highlights the need to \textit{delve deep and search wide} through development of small pathfinder experiments and new detector technologies~\cite{Battaglieri:2017aum, butler2023report}.

The dark photon is a hypothetical, low-mass vector boson, which has been posed as a candidate for dark matter.
Dark photons could account for much of the dark matter,
and are theoretically motivated via fluctuations of a vector field during the early inflation epoch of our universe.
A relic abundance of such a field could be produced
non-relativistically in the early universe through either the misalignment mechanism~\cite{PhysRevD.84.103501}
or through quantum fluctuations of the field during inflation~\cite{Graham:2015rva}. 

A dark photon is identified as the boson of an extra U(1) symmetry \cite{1986PhLB..166..196H}. Such a symmetry would mix between the Standard Model photon and the new gauge boson providing a detection portal.  The Lagrangian then varies from the Standard Model, $\mathcal{L}_{SM}$, as
\begin{equation}\label{eq:lagrange}
    \mathcal{L} = -\frac{1}{4} F'_{\mu\nu} F'^{\mu\nu} + \frac{1}{2} m^2 A'_\mu A'^\mu \\
    - \frac{1}{2}\epsilon F'_{\mu\nu} F_{EM}^{\mu\nu} + \mathcal{L}_{SM}.   
\end{equation}

\noindent
Here $m$ is the mass of the dark photon, $F_{EM}^{\mu\nu}$ and $A_\mu$ are the electromagnetic field tensor and gauge potential, $F'_{\mu\nu}$ and $A'_\mu$ are the dark photon field tensor and gauge potential, and $\epsilon$ is the dark photon-to-electromagnetic kinetic mixing factor which must be measured. 
The mixing term between the two coupled fields is then $\frac{1}{2}\epsilon F'_{\mu\nu} F_{EM}^{\mu\nu} $. Through kinetic mixing, dark photons would be detectable in electromagnetic searches, and $\epsilon$ can be measured. Null results place constraints on the mass-$\epsilon$ parameter space.

Astronomical observations \cite{Read:2014qva, 2019arXiv191014366D, de2021dark}, particularly the stellar data from the Gaia satellite, constrain the mean energy density of dark matter $\rho_{\text{DM}}$ at our position in our Galaxy: $\rho_{\text{DM}}$ $\approx\,$450 $\pm$100\,TeV/m$^3$. For dark photons in free space, converting this energy density to its corresponding electric field gives~\cite{Chaudhuri_2015}
\begin{equation}
\label{EobsVsEdark}
\bigl|\mathbf{E}_{\mathrm{ant}} \bigr| \approx \epsilon \sqrt{\frac{2}{\varepsilon_{0}}\rhoDM},  
\end{equation}
\noindent where $\mathbf{E}_{\mathrm{ant}}$ is an electric field observed at an antenna, $\epsilon$ is the small, dimensionless kinetic mixing parameter between the dark photon and electromagnetism and $\varepsilon_{0}$ is the permittivity of free space. This equation assumes that the dark photon corresponds to all of the dark matter. For our local dark matter energy density estimate Eq.\ \ref{EobsVsEdark} gives an electric field of $\bigl|\mathbf{E}_{\mathrm{ant}} \bigr| \approx \epsilon \times 4$ kV/m.


Knowledge of dark photon properties informs experimental design and search methodology. The small coupling to classical electromagnetism of order $\epsilon$ necessitates placing the experiment in an environment where the level of external interfering signals is significantly reduced. Dark photons, if they are a component of the dark matter, will be present everywhere including inside an RF-shielded room. The mass (or frequency) is unknown, motivating a wide search in frequency. The known stellar velocity dispersion in our Galaxy broadens the predicted linewidth to a quality factor $Q_\mathrm{DP} \approx 10^6$ \cite{Gramolin_2022}. This enables optimal filtering of the data for a narrowband, constant signal buried in the noise, as discussed below. 

At its core, the Dark E-Field Radio experiment is an FFT-based radio-frequency spectrum analyzer, searching for a small power excess on the wide band thermal noise spectrum received from an antenna in a cavity. Details are presented in Sec.\ \ref{sec:Experiment}. 

The remainder of the paper is arranged as follows. Section \ref{sec:Experiment} outlines the experimental details and how they are motivated by dark photon properties, as well as data collection. Section \ref{sec:analysis} provides details on how the data are processed to search for a dark photon signal, and provides a statistical exclusion limit on the lack of a signal detection. Section \ref{sec:calibration} takes the limit generated in Sec. \ref{sec:analysis} and converts it into a limit on $\epsilon$. Section ~\ref{sec:results} discusses systematic uncertainties, the elimination of a single candidate, and summarizes the null-result of our current run. Finally, Sec.\ \ref{sec:discussion} reviews our experiment and presents a plan for future searches. Table \ref{table:spectrum_labels} provides a summary of select parameters and references where they are first introduced. 

\section{Experiment}
\label{sec:Experiment}


\begin{figure}[ht]
\includegraphics[width=0.5\textwidth]{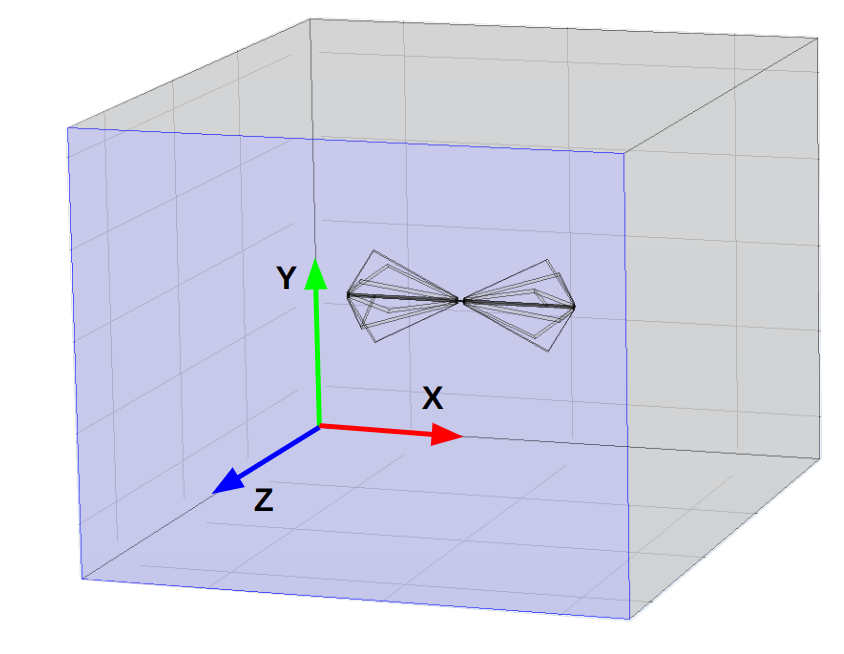}
  \caption{CAD model of the Dark E-Field Radio Experiment. Not shown is the low noise amplifier inside the room (see Fig.~\ref{fig:schematic} for further details). Nominal dimensions of the room are $3.05 \times 2.44 \times 3.66$\,m. The biconical dipole antenna is shown positioned in the center of the room with x-polarization. Polarization and position are changed multiple times during the course of the run, as described in 
  Sec.~\ref{sec:Experiment}.}
  \label{fig:room}
\end{figure}

The Dark E-Field Radio (DER) experiment consists of a linearly polarized biconical E-field antenna \cite{AB-900ADatasheet} (designed to operate between 25 and 300\,MHz) inside of a cavity (Fig.\ \ref{fig:room}). The cavity is a room-temperature, commercial, shielded room \cite{LindgrenDatasheet} which serves both to isolate the experiment from external radio frequency interference (RFI) and to provide resonant enhancement of potential dark photons after they have converted to Standard Model photons.



Figure 2 shows a schematic diagram of our radio-frequency spectrum analyzer system. A low noise amplifier (LNA), secondary amplifier chain and a band pass filter provide analog signal conditioning before the radio-frequency signal is directly digitized by a two channel ADC. From the ADC, time domain records are sent to a graphics processing unit (GPU) which performs a fast Fourier transform (FFT). This system is discussed further in Sec. \ref{sec:GPUbasedRTSA}.

An identical antenna is placed outside of the shielded room to monitor the local environment for RFI. If any candidate signal is detected by the main channel (B) and can be correlated within a 3 minute window to a similar signal in the "veto" channel (A), the contaminated scan can be excluded. This process is performed offline during the analysis stage. If no candidates are tagged in the analysis, these veto data are not used.

\begin{figure}
  \includegraphics[width=.48\textwidth]{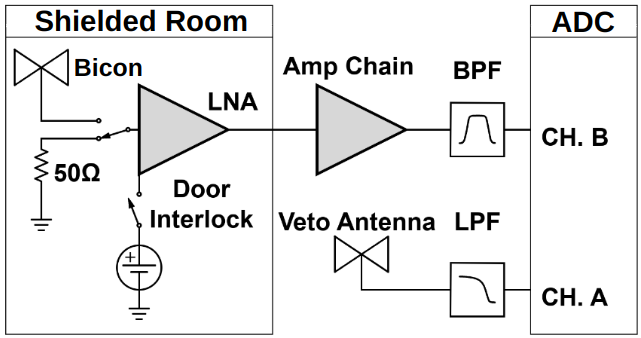}
  \caption{Schematic diagram of the Dark E-field Radio system with nominal gain and noise temperature of 74\,dB and 115\,K respectively. The shielded room provides $>\!110\,$dB of isolation over the 50-300\,MHz frequency span and radiates a room temperature black body spectrum which is the dominant background measured. An identical antenna outside of the shielded room monitors local radio frequency interference (RFI). The 50$\,\Omega$ terminator provides a known thermal noise source which is helpful for monitoring amplifiers over a data run and checking for interfering candidates. Band pass and low pass filters (BPF and LPF) provide anti-aliasing, and define the bandwidth of the signal received from both antennas.
  }
  \label{fig:schematic}
\end{figure}
\subsection{Experimental considerations}
\label{sec:experimentalRequirements}

\subsubsection{Background: Noise and Interfering Signals}
\label{sec:BackgroundNoiseOverview}
The overarching goal of the experiment is to measure weak, coherent, electric fields ($\approx$\,40\,pV/m)  embedded in a wide-bandwidth background. The background is primarily composed of coherent RFI, $\approx$\,100\,$\mu$V/m (measured in the lab), and room temperature black body spectrum$\,\approx$\,1\,nV/($\textrm{Hz}^{1/2}\,\textrm{m})$. To reduce the effects of RFI, the experiment is placed in an RF-shielded environment. In order to make external RFI a sub-dominant contribution to our total noise floor, shielding effectiveness (SE) in excess of 100\,dB is required. At the time of the experiment the SE was measured to be at least 110\,dB, and greater than  130\,dB at some frequencies. 

Contributions from the ADC (ADC Effects) further reduce our sensitivity. Introducing a gain factor $G(\nu)$ before the ADC greatly reduces these effects to near-negligible levels. The ADC's noise floor is approximately five orders of magnitude lower than the experiment's output-referred thermal noise density, making it negligible since this noise will average down along with the thermal noise of the experiment. Spurious signals produced internal to the ADC, however, will behave similarly to a candidate, emerging further above the noise with increased averaging. By terminating the input of the ADC, these spurious signals were measured. The most significant of these was $10^{-14}\,$W, output-referred. We correctly predicted this would become detectable after approximately 3 days of averaging. This is our only false positive signal and is discussed further in Sec. \ref{sec:falsePosSig}.        

The LNA contributes approximately 0.3\,nV/$\textrm{Hz}^{1/2}$ of input-referred noise (LNA Noise). Band-pass filters are added to reduce the total-integrated power received outside of the band-limited region of interest. This allows for more gain and therefore a smaller relative contribution of ADC Effects. 

The above considerations can be summarized in the following equation for output-referred power spectral density (PSD) $\mathrm{S}_\textbf{o}(\nu)$, assuming no dark photon signal is present:

\begin{equation}
\begin{split}
    \mathrm{S}_\textbf{o}(\nu) = \,  &\textrm{ADC Effects} \\ &+ G \, \Biggl(\biggl.\frac{\text{RFI}}{ \text{SE}} + \text{Thermal Noise}
    + \text{LNA Noise} \Biggr),
\end{split}
\end{equation}

\noindent
where each term is a function of frequency. We convert this \emph{output} referred PSD to an \emph{input} referred PSD by dividing by $G$; $\mathrm{S}_\textbf{i}(\nu) \equiv \mathrm{S}_\textbf{o}/G$. This quantity is standard in the literature, however we are more interested in the antenna referred spectrum $\mathrm{S}_\mathrm{ant}(\nu)$, which is the noise delivered to the LNA before the addition of LNA noise. 

\begin{align}
\label{eq:sysPow}
    \begin{split}
    \mathrm{S}_\mathrm{ant}(\nu) &\equiv \mathrm{S}_\textbf{i}-\textrm{LNA Noise}
    \\
    &= \frac{\textrm{ADC 
  Effects}}{G} + \frac{\textrm{RFI}}{ \textrm{SE}} + \textrm{Thermal Noise}
  \\
  &\approx \textrm{Thermal Noise} .
  \,\,\,
  \end{split}
\end{align}

$\mathrm{S}_\mathrm{ant}$ is a useful quantity because its only non-negligible component is the thermal noise received by the antenna which ultimately sets the sensitivity of the experiment (assuming negligible ADC Effects, RFI and no dark photon signal). It is worth noting that while $\mathrm{S}_\mathrm{ant}$ scales like the square root of total integration time, the limit on $\epsilon$ scales like the square root of $\mathrm{S}_\mathrm{ant}$, and therefore the quarter root of integration time. Our choice of a 9 day data run is set by this scaling (see \cite{GroupPaper} for a discussion).

\subsubsection{Frequency Resolution 
\texorpdfstring{$\DeltanuRF$}{△νRF}}
\label{sec:freqResolution}
As discussed in Sec. \ref{sec:intro} and in \cite{Gramolin_2022}, the fractional linewidth of the dark photon is relatively well
accepted. While we will use a specific Rayleigh lineshape  during data analysis, knowing the expected $Q$ of the dark photon line $Q_\mathrm{DP} \approx 10^6$, allows us to set the frequency resolution $\DeltanuRF$ of the FFT. Since our thermal background has a constant PSD, it seems advantageous to make $\DeltanuRF$ as narrow as possible in order to minimize the thermal noise power contained in each bin. However, once $\nu/\DeltanuRF > Q_\mathrm{DP}$, the signal's power will also be split between bins. While a narrow $\DeltanuRF$ splits both signal and noise between bins, decreasing $\DeltanuRF$ entails performing longer FFTs, which results in acquiring fewer FFTs for a fixed total integration time $\tau$. In other words, SNR improves as $\DeltanuRF$ is reduced until it reaches the expected width of the signal. For a constant $\tau$, it can be shown that
\begin{equation}
    \text{SNR}\propto
    \begin{cases}
   \DeltanuRF^{-1/2} & \text{if } \DeltanuRF \geq \nu/Q_\mathrm{DP} \\[3pt]
       \DeltanuRF^{1/2} & \text{if } \DeltanuRF < \nu/Q_\mathrm{DP}.
    \end{cases}
\end{equation}

\noindent This motivates choosing $\nu_\mathrm{min} / \DeltanuRF \approx {Q}_\mathrm{DP}$, or $\DeltanuRF \approx ~50$\,Hz, for a scan beginning at  $\nu_\mathrm{min}=50\,$MHz. $\DeltanuRF$ is set through the length of the FFT and sample rate. Constraints on these parameters dictate that we set $\DeltanuRF$ to 47.7\,Hz. 

\subsubsection{Clock stability}
\label{sec:clockStability}

The raw, digitized output of our experiment is a time series, sampled at the digitizer's clock rate. Since the discrete Fourier transform (DFT) of a perfect sinusoid sampled by an unstable clock will have a finite spectral width, clock stability must be better than the expected spectral width of candidate signals, which in our case is is set by the expected $Q\,\approx 10^6$. To achieve the required stability, we synchronize the sample clock  (Valon 5009 RF synthesizer) of our ADC to a $10\,$MHz rubidium frequency standard (Stanford Research Systems FS725) which is further steered by the one pulse-per-second (pps) signal from a GPS receiver. This system has medium and long term fractional frequency stability (Allan deviation) of $\sigma_y(\tau)<3\times10^{-12}$ (where $\tau$ is the averaging time) and phase noise of less than -65\,dBc/Hz at offset frequencies $>50$\,Hz from the carrier. This means that over the course of a single acquisition, the power contained in a bin will spread to an adjacent bin by less than 1 part in $10^6$ which is more than sufficient for our experiment.

\subsubsection{Statistical uniformity}
In practice, an antenna in a cavity exhibits extreme sensitivity to the location of any conductors within the cavity including the location of the antenna itself. The cavity and everything within it form a coupled resonant system whose resonances are strongly affected by the physical setup. Very small physical disturbances of the system result in large variations in measurements. To combat this effect and to allow for repeatable measurements, the concept of \textit{statistical uniformity} is used by those who employ cavities as \textit{reverberation chambers}\cite{NBSTechNote1506, NBSTechNote1508}. With this in mind, a necessary modification to the antenna aperture (defined in Sec. \ref{sec:calibration}) is to replace the notion of an antenna having a single aperture with that of an antenna-cavity system having an \textit{average} aperture, where the average is taken over many configurations of the system. We denote this averaging over system configurations with the use of $\langle \, \rangle$.\cite{NBSTechnote1092, TaiEffectiveApertureOriginal}

As previously stated, the resonances of the coupled antenna-cavity system are highly dependent on the configuration of the system. Each observed resonance occurs with some characteristics (i.e. frequency and $Q$) as a result of nearby modes interacting with it. By intentionally changing the configuration of the system, each individual mode will be pulled around by its neighbors. 
The configurations which are averaged over can be created through the rotation of metal mode stirrers which are large enough to occupy a significant volume of the cavity \cite{ReverbChambersALaCarte}. 

We accomplish a similar effect by moving the antenna to 9 positions and polarizations throughout the course of the run. This is similar to the effect of a turntable frequently employed by microwave ovens. While this does not ensure statistical uniformity, it allows us to perform a simulation which we find agrees reasonably well with measurement. This is further discussed in Sec. \ref{sec:ResultsUncertaincities}.  

Proper mode stirring requires the reverberation chamber to be operated where the modal density is sufficiently high, far above the so-called lowest usable frequency. This is approximately 200\,MHz for a cavity of our dimensions \cite{hill2009}. Operating in this regime allows for an approximate analytic solution of chamber parameters which is outlined in \cite{hillPlaneWaves}. Since we operate the experiment at frequencies below 200 MHz (the \textit{undermoded region}), we do not have a well stirred chamber and must rely on simulation for calibration. We plan to upgrade our chamber with mode stirrers in future, higher frequency runs.

\subsection{GPU-based real-time spectrum analyzer}
\label{sec:GPUbasedRTSA}

The use of commercial Spectrum Analyzers (SAs) which feature so-called real time spectrum analyzer (RTSA) mode come with several restrictions which limit the efficiency with which they are able to perform wide-band scans with narrow frequency resolution, as we pointed out in our pilot run \cite{GroupPaper}. The number of frequency bins output by a real discrete Fourier transform (DFT) is equal to half of the number of time domain samples, while the bandwidth is given by half of the sample rate. Furthermore, the ability to acquire data in real time requires a DFT algorithm (generally implemented as a fast Fourier transform, FFT) and computational resources which can operate on time domain data at least as fast as it is acquired. In practice, real-time DFTs with high frequency resolution and wide bandwidth require modest memory, transfer rates and processing resources. For this reason, we have constructed our own SA based on the Teledyne ADQ32 PCIE digitizer, which is wide-bandwidth (up to 1.25\,GHz), high resolution ($2^{24}$ point FFT), and nearly 100\% real-time. Indeed, we have been unable to find a commercial SA with comparable capabilities.

The first component of our SA is the Teledyne ADC which directly samples at a rate of $800\,$MHz and bit depth of 12 bits. From the ADC, records of length $2^{24}$ are acquired and sent to a graphics processing unit (GPU) which computes an FFT. Approximately $10^4$ FFTs are performed and added to a cumulative sum on the GPU (representing about 3 minutes of real time data). Dividing by the number of FFTs provides an averaged spectrum that is saved for offline processing. This \textit{pre-averaging} reduces the raw $\approx$~1.5\,GB/s/channel data stream to $\approx$~\ 0.15\,MB/s/channel, which greatly reduces storage requirements. However, this comes at the cost of temporal resolution of transient candidates.

\subsection{Data acquisition run}

\begin{figure}[ht]
\includegraphics[width=0.4\textwidth]{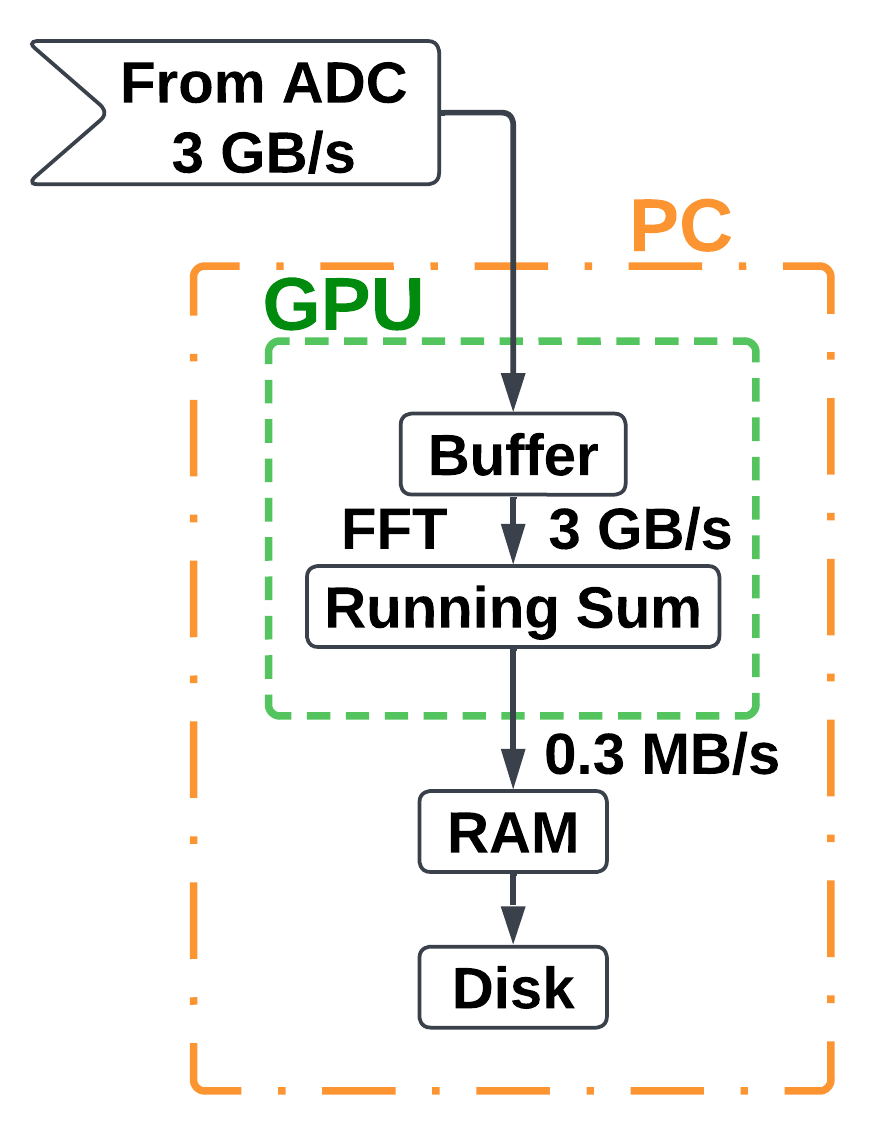}
  \caption{Real time DAQ data stream. Approximately $10^4$ time series records (about 3 minutes of real time data) are written from the ADC directly to GPU buffers. FFTs are performed on these records resulting in a pre-averaged spectrum which can be saved to disk. This set up is duplicated for channels A and B, though the data rates indicate the sum of both channels. }
  \label{fig:dataProcessing}
\end{figure}

Data were collected during a 9-day run from May 10 to May 19, 2023. Each day was subdivided into data-collection (23 hours 15 minutes) and setup (45 minutes) periods. The setup period includes moving the antenna, changing a 12\,V battery for the LNA, file management and documentation. In order to reduce the data rate and storage requirements, all data were pre-averaged into 3-minute chunks and then saved. This is shown in Fig. \ref{fig:dataProcessing}. For the data analysis, all 9 days of data were averaged together to create a single spectrum $\mathrm{S}_{\textbf{o}}$ (Fig. \ref{fig:inputOutputPowSpec}). If candidates are found, their time dependence can be observed by looking at the 3-minute pre-averages. All further analysis is performed on the full 9-day $\mathrm{S}_{\textbf{o}}$ spectrum and is described below (Sec. \ref{sec:analysis}).

\subsection{Raw data, \texorpdfstring{$\mathrm{S}_\textbf{o}$}{So}}
\label{sec:rawDataSo}
\begin{figure}
  \includegraphics[width=0.48\textwidth]{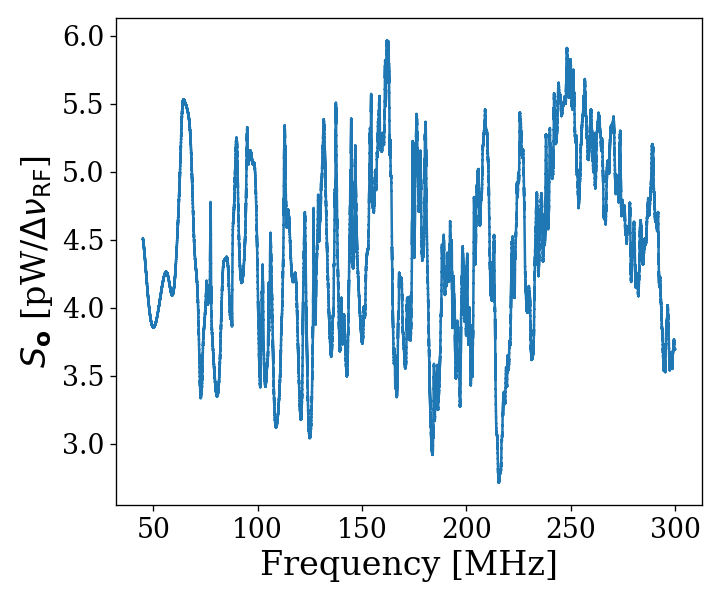}
  \caption{Averaged output-referred power spectrum, $\mathrm{S}_\textbf{o}$. Data were taken over a 9 day period at 9 antenna positions. The narrow variations are mainly due to the effective temperature difference between the room and LNA, though there is a small contribution due to amplifier gain and noise temperature variations (see Sec. \ref{sec:rawDataSo}). The variations seen here are not noise; their shape is repeatable for a given antenna position. The noise on this background is not visible at this level of zoom, but can be seen in Fig. \ref{fig:analysisA}, which shows a zoomed-in view of the spectrum at $240\,$MHz.}
  \label{fig:inputOutputPowSpec}
\end{figure}

All 9 days of pre-averaged data from the run are averaged together. The stability of our sample clock ensures that this is a simple process; frequency bins corresponding to a given frequency are added and normalized to the total number of pre-averaged spectra. This process produces the raw spectrum, $\mathrm{S}_\textbf{o}$ (Fig. \ref{fig:inputOutputPowSpec}), on which we will perform a search for power excess.

Inspection of $\mathrm{S}_\textbf{o}$ reveals small power variations over spans of tens of kHz. Given an antenna in a cavity in thermal equilibrium with the input of an LNA, whose input is assumed to be real and matched, one would expect an output PSD which is constant with respect to frequency up to small variations in LNA gain. The theory for this is outlined in \cite{Dicke:1946}. These variations are not noise; for a given antenna position we repeatedly measure the same shape (though the noise riding on these variations \emph{is} random). The origin of the observed small variations lies in the effective temperature difference between the room and LNA causing a net power flow from the antenna into the LNA. This effective temperature difference partially excites modes of the antenna/cavity system, causing the observed variations. We suspect this effect originates from a small reactive component of the LNA's input causing the electronic cooling described originally by Radeka \cite{electronicCooling}. This effect can be eliminated by adding an isolator between the antenna and LNA \cite{haystacDesign, SRFC} (though for our wide-band search this introduces more complications than it solves so we omit it).  

\section{Data Analysis}
\label{sec:analysis}

\begin{figure}[ht]
\includegraphics[width=0.48\textwidth]{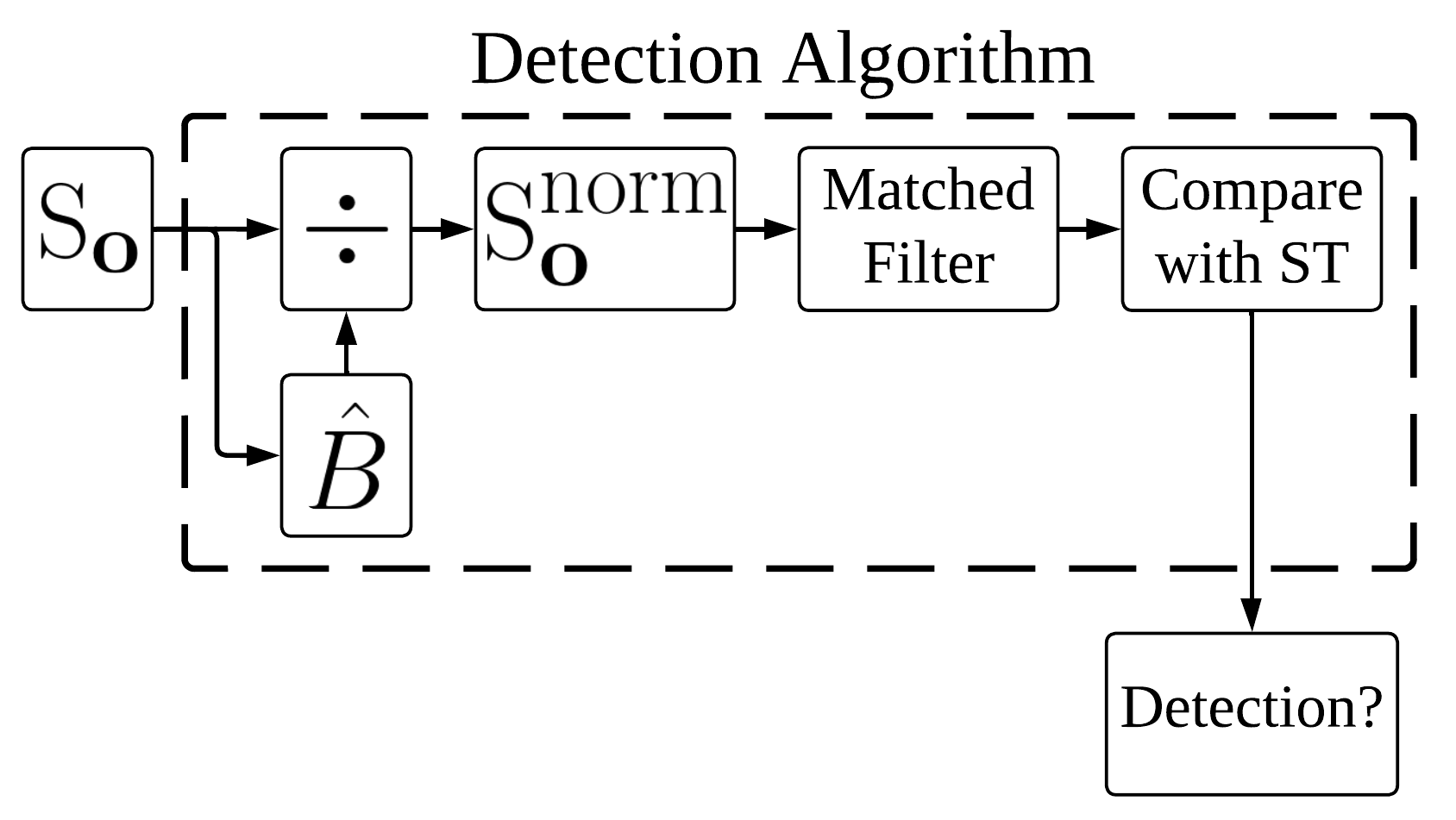}
  \caption{Flow chart outlining the logic of signal processing in the detection algorithm of sections \ref{sec:FitB} through \ref{sec:matchedFilter}. $\hat{B}$ is the smoothed fit to $\mathrm{S}_\textbf{o}$ generated by low pass filtering. The output, \textit{Detection?}, is a Boolean array which signifies a detection or lack thereof at each frequency bin. We detect a candidate if a bin contains more power than a significance threshold (ST) (Sec. \ref{sec:normalizedSpectrum}).}
\label{fig:detectorFlowChart}
\end{figure}

At this point, we have compiled a single, averaged, output-referred power spectrum, $\mathrm{S}_\textbf{o}$ (Fig. \ref{fig:inputOutputPowSpec}). The task of \textit{analysis} is to extract a dark photon signal from this spectrum if it is present. Otherwise, in its absence, we would like to set a limit on the amount of output-referred power one would be able to detect \emph{most of the time} were a narrow signal to be present in this averaged dataset. We quantify the meaning of "most of the time" by conducting a series of Monte Carlo "pseudo-experiments" on artificial signal-containing spectra for synthetic signals of varying powers and frequencies. The following subsections are organized as follows: 

\bigskip
\bigskip
\begin{enumerate}
    \item[\ref{sec:FitB}:] Fit $\mathrm{S}_\textbf{o}$ to extract an estimate of the background $B$ (which we call $\hat{B})$ whose origin was discussed in Sec. \ref{sec:rawDataSo}. See Fig. \ref{fig:analysisA}.
    \item[\ref{sec:normalizedSpectrum}] Divide the spectrum by $\hat{B}$ to generate the \emph{normalized spectrum}, which very nearly follows a Gaussian distribution. Discuss statistics of the normalized spectrum and choose a global significance level and its associated \textit{significance threshold}. See Fig. \ref{fig:analysisB}.
    \item[\ref{sec:matchedFilter}] Apply a matched filter to the normalized spectrum and establish a significance threshold on its output using the same method defined in the previous section. See Fig. \ref{fig:analysisC}. The previous three steps comprise our \emph{detection algorithm} which is shown in Fig.~\ref{fig:detectorFlowChart}.
    \item[\ref{sec:monteCarlo}] Perform a Monte Carlo analysis to simulate the required power of a signal that can be detected above the significance threshold 95\% of the time. We use this to recover a 95\% exclusion limit on the output referred power spectrum.
\end{enumerate}

In Sec. \ref{sec:calibration} we convert this threshold on $\mathrm{S}_\textbf{o}$ into an actual limit on $\epsilon$.

Throughout the figures of this section we will follow a relatively large ($40\,$fW, output-referred) software-injected, synthetic dark photon signal at 240$\,$MHz to illustrate what a candidate would look like as it passes through the analysis procedure. This signal is added to $\mathrm{S}_\textbf{o}$. For clarity, we have removed a single interfering candidate which we discuss in Sec. \ref{sec:falsePosSig}.

\subsection{Fit background, \texorpdfstring{$\boldsymbol{\hat{B}(\nu)}$}{Bv}}
\label{sec:FitB}
As shown in Fig. \ref{fig:inputOutputPowSpec}, the measured power spectrum looks like flat thermal noise \emph{multiplied} by some frequency dependent background, $B(\nu)$. However, for this section we will not concern ourselves with the origin of $B$ or any details of the experiment aside from two assumptions:

\begin{enumerate}
    \item The measured background is the product of a normally distributed spectrum and some background. This is enforced by the central limit theorem due to the large number of averaged spectra, independent of any experimental specifics. 
    \item The line shape of the signal is known and the width of this signal is much narrower than the width of features on the background, viz. $\Delta \nu_\mathrm{DP} << \Delta \nu_{\mathrm{B}}$
 \end{enumerate}

The first assumption (1) implies that if we were able to extract the background, dividing $\mathrm{S}_\textbf{o}$ by this extracted background would yield a \emph{dimensionless}, normally distributed power spectral density on which to perform a search for a dimensionless signal. The second assumption (2) will be critical in both performing the fit to the background (this section), and performing matched filtering (Sec. \ref{sec:matchedFilter}).

In light of these assumptions, we attempt to fit for the background power spectrum. Since this fit estimates $B$, we use the symbol $\hat{B}$ to refer to it. As discussed in \cite{haystac_2017}, a particularly effective fitting technique that can discriminate between the wide bumps of $\mathrm{S}_\textbf{o}$ and a narrow signal is to use a low pass filter. We implement this filter in two stages:

\begin{enumerate}
    \item A median pre-filter (51 bins or about 2.4\,kHz wide) attenuates any very narrow, very large excursions which would interfere with any following filters, causing them to "ring"
    \item A 6\textsuperscript{th}-order Butterworth low pass filter 
 (corner frequency of 210 bins or 10\,kHz)
\end{enumerate}

\begin{figure}[H]
  \includegraphics[width=.48 \textwidth]{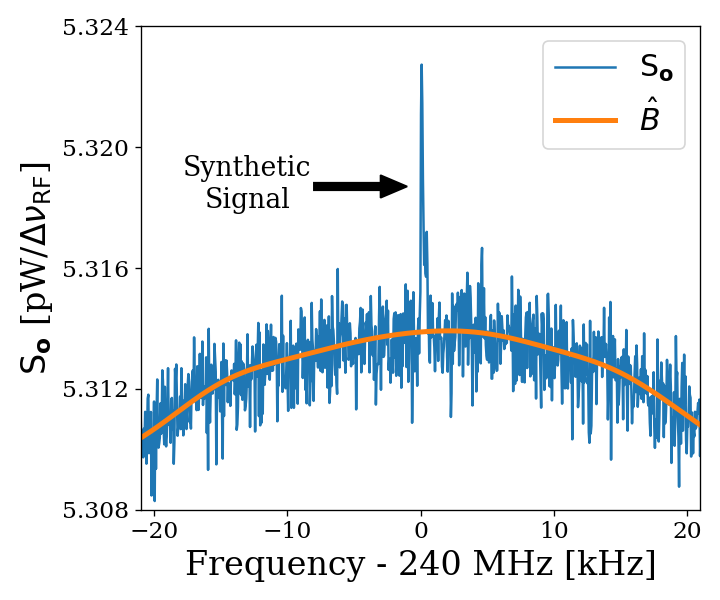}
\caption{Fitting background $\hat{B}$ in the presence of a synthetic signal injected at 240\,MHz. Starting from the averaged, output-referred spectrum ($\mathrm{S}_\textbf{o}$), we fit the background using a series of filters (section \ref{sec:FitB}, and Fig. \ref{fig:detectorFlowChart}). This figure is a highly zoomed in view (240\,MHz $\pm$ 20\,kHz) in order to show the noisy Rayleigh signal shape. }
\label{fig:analysisA}
\end{figure}

These bin widths/frequencies should be interpreted as the width of spectral features on $\mathrm{S}_{\textbf{o}}$ that are attenuated and will, therefore, not show up in the background fit. A narrow zoom of this fit with a synthetic signal is shown in orange in Fig. \ref{fig:analysisA}. 

\subsection{Normalized spectrum, \texorpdfstring{$\boldsymbol{\mathrm{S}}_\textbf{o}^\mathrm{norm}$}{sNorm}}
\label{sec:normalizedSpectrum}
\begin{figure}[ht]  \includegraphics[width=0.48\textwidth]{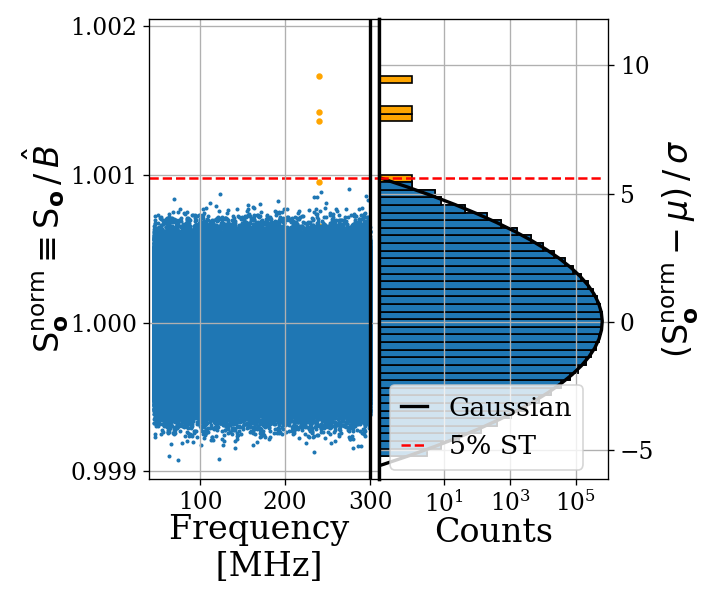}
  \caption{Dividing $\mathrm{S}_\textbf{o}$ by $\hat{B}$ yields a dimensionless, normally distributed power spectrum that we define as $\mathrm{S}_\textbf{o}^{\mathrm{norm}}$. We show $\mathrm{S}_\textbf{o}^{\mathrm{norm}}$ in two ways: a normalized power/frequency spectrum (\textit{left}) and rescaled into Z-score units and collapsed into a histogram (\textit{right}). The histogram shows power excess and Gaussian fit, but frequency information is lost. We compute a 5\% significance threshold ST (\textit{dashed red}), above which we will detect a candidate by chance 5\% of the time. Bins adjacent to the 240MHz synthetic signal show up in orange on both plots. A single interfering signal has been removed for clarity. We discuss this further in Sec. \ref{sec:falsePosSig}}
  \label{fig:analysisB}
\end{figure}

Once we have a fit to the background, $\hat{B}$, division of $\mathrm{S}_\textbf{o}$ by this fit yields a dimensionless, Gaussian distributed spectrum  
\begin{equation}
    \mathrm{S}_\textbf{o}^\mathrm{norm} \equiv \frac{\mathrm{S}_\textbf{o}}{\hat{B}}.
\end{equation}
As we discuss in \cite{GroupPaper}, this normalized spectrum (Fig. \ref{fig:analysisB}) should have a mean $\mu_{\mathrm{norm}}$=1 and a standard deviation given by the Dicke radiometer equation $\sigma_{\mathrm{norm}} =(\tau \DeltanuRF)^{-1/2}$  where $\tau$ is the total integration time ($\approx ~\,$9 days) and $\DeltanuRF$ is the width of a bin (47.7\,Hz). This works out to a predicted  $\sigma_{norm}$ of $1.727 \times 10^{-4}$. $\mu_{\mathrm{norm}}$ and $\sigma_{\mathrm{norm}}$ calculated from the data are $1-1.2 \times10^{-5}$ and $1.741\times10^{-4}$ respectively, which agree with the predicted values to better than 1\%. Knowing the statistics of the background allow us to set a threshold above which we have some confidence that a candidate is not a random fluctuation. We briefly derive this threshold as described below.

The probability that all N bins are less than \textit{z} standard deviations, $z\sigma$, for a standard normal distribution is given by $\Bigl\{\frac{1}{2}\left[1 + \textrm{erf}\left(z/\sqrt{2}\right) \right]\Bigr\}^N$, where $\textrm{erf}(z)$ is the standard error function and \textit{z} is real. Setting this equal to $100\%-$X (where X is the \textit{significance} or the desired probability a fluctuation crosses the $z\,\sigma$ threshold assuming no signal), and inverting $\textrm{erf}(z)$ yields a significance threshold (ST). We choose X = 5\% corresponding to a 5\% probability that an observed fluctuation above this ST is due to chance rather than a significant effect (i.e. a signal). A 5\% ST for $5.2\times 10^{6}$ bins (our 50-300\,MHz analysis span) works out to 5.6$\sigma$. This is shown in Fig.~\ref{fig:analysisB}. 

It should be noted that it is common in physics to discuss "5$\,\sigma$ significance". This means that a given experiment has a 1\,-\,erf($5/\sqrt{2}$) probability (about 1 in $3\times 10^{6}$)  of a false positive. The analysis of our normalized spectrum involves testing many independent frequency bins to see if any one of them exceeds some threshold. It is helpful to view these bins as "independent experiments", each involving a random draw from the same parent Gaussian distribution. In this context, we discuss global significance (all of the bins) in contrast to local significance (a single bin). Setting a global 5\% significance threshold is equivalent to setting a local threshold of 5.6$\,\sigma$ given $5.2\times 10^{6}$ bins.

It is possible to set a simple limit using this significance threshold on the normalized spectrum, which was our method in \cite{GroupPaper}. However, knowing the line shape of the dark photon signal provides additional information that improves sensitivity (up to a factor of $\approx2$) at the higher frequency end of the spectrum, as shown in Fig. \ref{fig:outputLim}.

\subsection{Matched filter}
\label{sec:matchedFilter}
As discussed in \ref{sec:normalizedSpectrum}, one simple method to set a limit is to look for single-bin excursions above some threshold. However, galactic dynamics impart a dark photon candidate with a Rayleigh-distributed, spectral signature, which has a dimensionless width $Q_\mathrm{DP}\approx 10^6$ \cite{Gramolin_2022}. This means that the expected width of a candidate signal over our analysis span (50-300\,MHz) varies between 50-300\,Hz. We set $\DeltanuRF$ = 47.7\,Hz to maximize SNR for the lowest expected signal width (see Sec. \ref{sec:freqResolution}). However, this divides signal power between adjacent bins, an effect that becomes more pronounced at higher frequencies, leading to a decrease in sensitivity. By using a signal processing technique known as \textit{matched filtering} \cite{MatchedFilterIntro}, we restore some of the sensitivity lost due to the splitting of signal between the fixed-width frequency bins of the FFT.

\begin{figure}[ht] \includegraphics[width=0.48\textwidth]{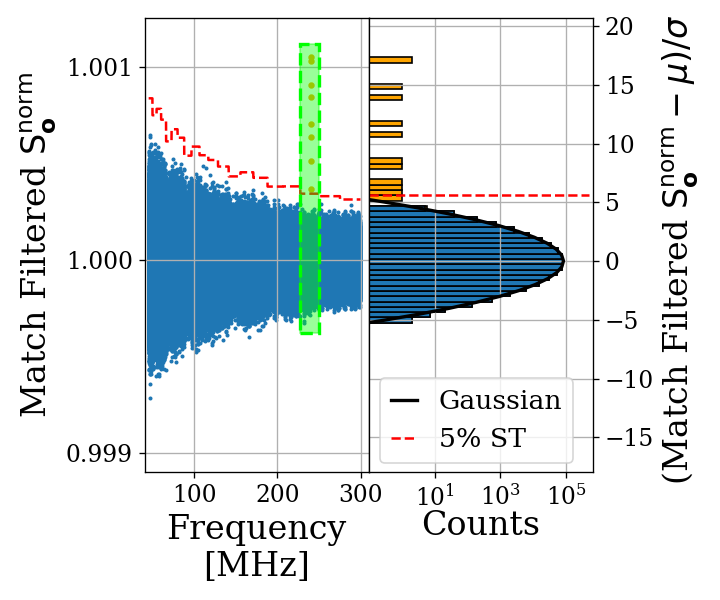}
  \caption{$\mathrm{S}_{\textbf{o}}^{\mathrm{norm}}$ after it has been passed through a matched filter. The template varies in width throughout the frequency span resulting in 20 subspans, each with a constant 5\% significance threshold ST (\textit{dashed red}). Histogram only includes 227-250\,MHz subspan (enclosed in the green box). The signal-to-threshold ratio of the synthetic signal (orange) improves by a factor of about 1.8 as compared to Fig.~\ref{fig:analysisB} without a matched filter. The frequency dependence of this effect is shown in Fig. \ref{fig:outputLim}. A single interfering signal has been removed for clarity.}
  \label{fig:analysisC}
\end{figure}

For a known signal shape, the detection technique which optimizes SNR is the matched filter. This is implemented on the normalized power spectrum using the Rayleigh spectral line shape of \cite{Gramolin_2022} as a template. Since we have a constant $\DeltanuRF$ and expect the width of the signal to vary across our span, we must calculate several templates of varying width to match the expected line shape. Every 10\% of fractional frequency change, a new template is generated and used to search that small sub-span of the normalized spectrum, each of which is also normally distributed though with its own standard deviation. This results in 20 subspans (50-55\,MHz, 55-60.5\,MHz etc.). The normalized spectra of all 20 subspans and the histogram of the 227-250\,MHz subspan are shown in Fig. \ref{fig:analysisC}.

As the width of the templates increase, the standard deviation of the output decreases, resulting in the $\nu^{-1/2}$ shape of the 5\% significance threshold shown in Fig. \ref{fig:analysisC}. It should be noted that since the total number of bins remains 5.2 million, the 5\% significance threshold still corresponds to 5.6$\sigma$; the shaping in Fig. \ref{fig:analysisC} is due to the variation in $\sigma$ for different templates, not a change in the $z= 5.6$ pre-factor.

\subsection{Monte carlo: pseudo experiments}
\label{sec:monteCarlo}
The previous three sub-sections outline the procedure for detecting the presence of a signal of known spectral line shape embedded in wide-band noise. We refer to this procedure as a \emph{detection algorithm} (see Fig.~\ref{fig:detectorFlowChart}) which we now calibrate through a Monte Carlo method. 

A synthetic spectrum is constructed by multiplying some $B$ by randomly generated Gaussian white noise characterized by $\mu_\mathrm{norm}$ and $\sigma_\mathrm{norm}$, as discussed in section \ref{sec:normalizedSpectrum}. A signal of known, total integrated, output-referred power and frequency, $\lambda(p, \nu)$, can now be added to this spectrum to create a test spectrum which can be passed through the detection algorithm. The frequencies of the synthetic signals are evenly spaced (approximately every $10\,$MHz). However because the signal spans a limited number of bins (one to six), the shape of the discretized signal is very sensitive to where its peak lands relative to the bins. To compensate for the fact we don't know where a dark photon's peak would land relative to the frequency bins, the frequency of the synthetic signal is randomly jittered by $\pm \DeltanuRF/2$, which is drawn from a uniform probability distribution at each iteration of the Monte Carlo. By repeating this with randomly generated Gaussian noise and various synthetic signals (including a small jittering of signal frequency outlined above), statistics are built up about how much total integrated power is required to detect a signal as a function of frequency \emph{most of the time}. We quantify this as the statistical power of the detection algorithm and denote it $100\% - \mathrm{Y} = 95\%$ following the standard convention of hypothesis testing.

\begin{table}[ht]
\begin{tabular}{|c|c|c|}
\hline
 & Only Noise & Noise + Signal \\
\hline
Detection & X  &  \( 100\% - \mathrm{Y} \)  \\
\hline
No Detection &  \( 100\% - \mathrm{X}\) &  \ Y  \\
\hline

\end{tabular}
\caption{Threshold parameters that are part of the detection algorithm and Monte Carlo. X is the significance of the analysis. It is a parameter passed to the detection algorithm which specifies the significance threshold. The quantity $100\% - \mathrm{Y}$ is the statistical power of the analysis. It is a parameter in the MC, which specifies a threshold on signal power where a given signal is detected in $100\% - \mathrm{Y}$ of the MC iterations. We choose both X and Y = 5\%.}
\end{table}

\begin{figure}[ht] \includegraphics[width=0.48\textwidth]{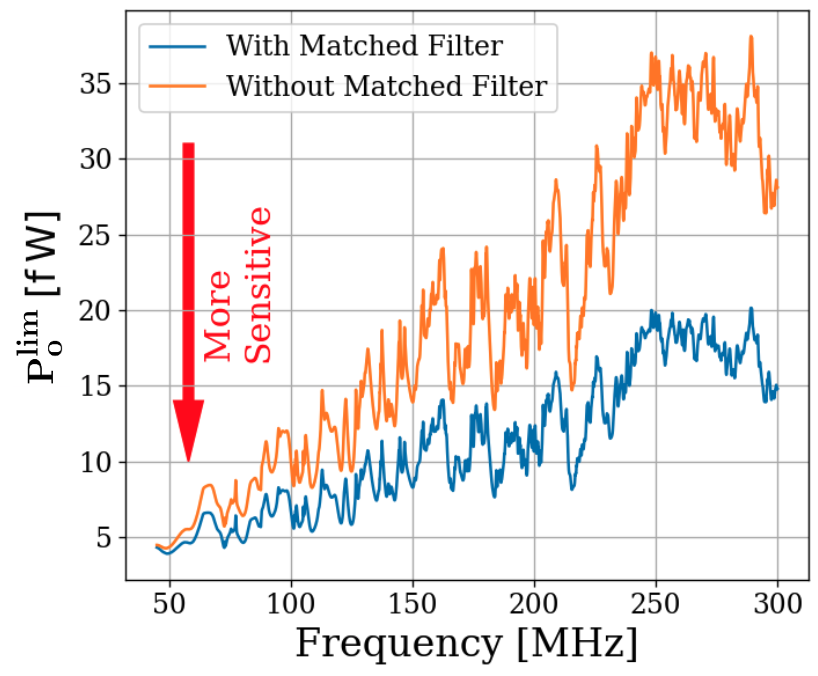}
  \caption{Limit on output-referred total integrated signal power, $\mathrm{P}_\mathbf{o}^\mathrm{lim}$. Limits computed with (\textit{blue}) and without (\textit{orange}) a matched filter (Sec. \ref{sec:matchedFilter}). The limits are similar at lower frequency but the matched filter improves sensitivity at higher frequencies where the signal power is split among more bins. The blue curve is used in the following sections.}
  \label{fig:outputLim}
\end{figure}

This Monte Carlo allows us to treat the detection algorithm as a black box which can be calibrated by passing it a known input (a synthetic $\mathrm{S}_{\textbf{o}}$ containing a synthetic signal, both software-generated) and looking at its output; a Boolean array of frequency bins representing signal detection. The limit on the total power contained in injected signals which can be detected 95\% of the time $\mathrm{P}_{\textbf{o}}^\mathrm{lim}$, is shown in Fig. \ref{fig:outputLim} in blue. Also shown in Fig. \ref{fig:outputLim} is a limit that does not include any matched filtering (orange) to highlight the frequency dependent improvement of the matched filter; this limit is only for illustration and not used in the following sections.

The limit set in this section is referred to the output of the amplifier chain. The topic of the next section will be to work back through the amp chain, to an E-field limit in the cavity and ultimately to a limit on $\epsilon$.

\section{Calibration of \texorpdfstring{$\boldsymbol{\epsilon}$}{e} limit}
\label{sec:calibration}
In this section we describe the calibration of our experiment and estimate our uncertainty. The previous section concluded with a limit on the output-referred power $\mathrm{P}_\textbf{o}^\mathrm{lim}$ (Fig. \ref{fig:outputLim}), which we now must convert into a frequency dependent limit on $\epsilon$.

We begin by inverting Eq. \ref{EobsVsEdark} 
\begin{equation}
\label{eq:epsLimEfield}
    \epsilon(\nu) < \sqrt{\frac{ \bigl|\mathbf{E}_\textrm{ant}^{\mathrm{lim}}\bigr|^2 \, \varepsilon_0}{2 \, \rho_{DM}}}\, ,
\end{equation}

 \noindent where the $lim$ superscript indicates a limit, below which a detectable electric field may be hiding. The $<$ should be taken to mean that in setting a limit on $\bigl|\mathbf{E}_\textrm{ant}^{\mathrm{lim}}\bigr|$,  $\epsilon$ is constrained to be less than the right hand side (if it exists at all).

 As discussed in Sec. \ref{sec:experimentalRequirements}, the first step of calibration is to convert from output-referred power to \textit{antenna referred power}. This represents the signal power presented to the LNA by the antenna via a matched transmission line and is given by

 \begin{equation}
\mathrm{P}_\mathrm{ant}(\nu) = \frac{\mathrm{P}_\mathrm{\textbf{o}}}{G} - T_\mathrm{amp} \, \kB \, \DeltanuRF ,    
\end{equation}

\noindent where $G$ and $T_\mathrm{amp}$ are the frequency-dependent amplifier gain and noise temperature ($74 \mathrm{-} 75\,$dB and $100\mathrm{-}120\,\mathrm{K}$ respectively, measured via the Y-factor method \cite{yfactorRhodeAndSchwartz}) and $\kB$ is Boltzmann's constant. 

Ultimately, the exclusion limit is set by fluctuations on this baseline described by
 
 \begin{equation}
 \label{eq:pLimAnt}
 \begin{split}
\mathrm{P}_\mathrm{ant}^\mathrm{lim}(\nu) &= \frac{\mathrm{P}_\mathrm{\textbf{o}}^\mathrm{lim}}{G} - \left(\frac{2}{n}\right)^{1/2}T_\mathrm{amp} \, \kB \, \DeltanuRF \\
&= \frac{\mathrm{P}_\mathrm{\textbf{o}}^\mathrm{lim}}{G} - \left(\frac{2\,\DeltanuRF}{\tau
}\right)^{1/2}T_\mathrm{amp} \, \kB,
\end{split}
\end{equation}

\noindent where the \textit{lim} superscript indicates an exclusion limit, \textit{n} is the total number of spectra averaged together, and $\tau$ is the total integration time. In the second line we have used $n =\DeltanuRF \, \tau$. In practice, the LNA correction is small; the first term divided by the second varies with frequency between 7 and 50. The $\tau^{-1/2}$ dependence of $\mathrm{P}_\mathrm{\textbf{o}}^\mathrm{lim}$ is implicit because it was calculated from $\mathrm{S}_\textbf{o}$ which is itself an averaged spectrum. As mentioned above, this $\tau^{-1/2}$ dependence implies that the limit on epsilon scales as $\tau^{-1/4}$.  

In the remainder of this section we explore the relationship between $\mathrm{P}_\mathrm{ant}^\mathrm{lim}$ and $\bigl|\mathbf{E}_\textrm{ant}^{\mathrm{lim}}\bigr|$ allowing us to use our experimental data to set a constraining limit on $\epsilon$ by employing Eq. \ref{eq:epsLimEfield}.

\subsection{Average effective aperture, \texorpdfstring{$\boldsymbol{\langle A_{e}(\nu)\rangle}$}{<Ae>(ν)}}
\label{sec:avgEffAperture}

An antenna's effective aperture, $A_e \, [\mathrm{m}^2$], represents the effective area that it has to collect power density or irradiance [$\mathrm{W/m}^2$] from an incident Poynting vector. It is a directional quantity which varies with the antenna's directivity $D(\Omega)$, where $\Omega$ represents solid angle around the antenna. It varies with frequency $\nu$, though it is generally discussed in terms of wavelength $\lambda$. Three matching parameters are introduced to describe how much actual power the antenna is able to deliver to a transmission line; $p$ the polarization match of the wave to the antenna, $m$ the impedance match of the antenna to the transmission line and $\eta_a$ the efficiency of the antenna which represents how much power is absorbed compared to that lost to Joule heating of the antenna. $p$, $m$ and $\eta_a$ are all real, dimensionless and vary between 0 and 1.

\begin{equation}
\label{eq:effAppFreeSpace}
    A_{e} \equiv \frac{\lambda^2}{4 \pi} D(\Omega) p\,m\,\eta_a .
\end{equation}

We define $A_{e}$ following \cite{hillPlaneWaves}, though some authors do not include $p$ in the definition \cite{Stutzman02} \cite{OriginalKraus}. 

$A_{e}$ is useful for an antenna in free space, however some modifications must be made to construct an analogous quantity for an antenna in a cavity. 

The first modification is to average over many configurations of the system. The background for this is given in Sec. \ref{sec:Experiment}. As discussed, we denote this averaging with  $\langle \, \rangle$ so that the average, effective aperture is denoted  $\langle A_{e} \rangle$. It is interesting to note that by averaging over configurations (namely antenna direction), $\langle A_{e} \rangle$ simplifies since $\langle D(\Omega) \rangle = 1$ by construction \cite{hillPlaneWaves}.

The second modification is to introduce a resonant enhancement factor which corresponds to the system's tendency to "ring up" in the same way any resonator will. We refer to this as \textit{composite} $Q$ and represent it as $\widetilde{Q}$. It is analogous to the standard quality factor of a resonator with one important modification; we operate our experiment across a wide frequency range so we define $\widetilde{Q}$ across the continuum of these resonances, not only on classical eigenmodes of the system.

These modifications allow us to construct a relationship between an observable E-field ($\mathbf E_{\mathrm{ant}}$ in Eq.\ \ref{EobsVsEdark} and \ref{eq:epsLimEfield}) and the power available at the port of an antenna for a given aperture

\begin{equation}
\label{eq:powerVsEobs}
    \langle \mathrm{P}_\mathrm{ant} \rangle = \frac{\lvert \mathbf{E}_{\mathrm{ant}}\rvert^2}{\eta_0} \langle \widetilde Q \, A_{e}\rangle,
\end{equation}

\noindent where $\eta_0$ is the impedance of free space. With this in mind, we perform an RF simulation to compute $\langle \widetilde Q \, A_{e}\rangle$.

\subsection{Simulation of \texorpdfstring{$\langle \widetilde Q \, A_e \rangle$}{<QAe>}}
\label{ssec:SimAe}
It is difficult to make claims about statistical uniformity in the ``undermoded'' regime where modes are not sufficiently mixed \cite{UndermodedRegime}, so we have employed a commercial electromagnetic finite-element modeling software package (COMSOL Multiphysics RF module \cite{COMSOLRef}). Within the simulation, a model of the antenna (with a 50$\,\Omega$ feed) is placed in a simplified room with wall features removed. Spot testing at various frequencies has shown that averaging results from various antenna positions using this simplified simulation behaves very similarly to one with the room features included at a fraction of computational complexity. 

Two similar simulations are run; driving an E-field while measuring the antenna's response and driving a second small monopole antenna and measuring the response of the primary antenna.

In the first simulation, we drive currents on the walls which correspond to a surface E-field magnitude of $1\,\mathrm{V/m}$ (made up of equal components in the x, y and z directions) using COMSOL's source electric field option. This field takes the place of $\mathbf{E}_{\mathrm{ant}}$ in Eq.~\ref{eq:powerVsEobs}. The antenna/cavity system resonates and causes an enhancement by $\widetilde{Q}$. The power received at the antenna's port is measured, allowing the calculation of $\widetilde Q \, A_e$, again from Eq.~\ref{eq:powerVsEobs}. By repeating this simulation for several positions, averaging allows us to compute $\langle \widetilde Q \, A_e \rangle$.

The second simulation shares the same geometry, but is used to compute a correction factor to account for differences between simulation and measurement and to estimate uncertainty on the first simulation through comparison to physical measurement. Rather than driving the system through currents on the walls, power is injected into the system with a 40\,cm monopole. From this simulation, two port scattering parameters (S parameters, defined in \ref{sec:simulationUncertainty}) are computed. A similar test is performed on the physical system using a vector network analyzer (VNA) which provides a physical measurement of the S parameters to compare with the simulation. The processing of the simulated and measured S parameter datasets are discussed in the following sub-section.

Both simulations are run at the same 18 positions; 9 of which are are approximately equivalent to the physical antenna positions while the other 9 are different in order to estimate how many positions are required for decent convergence of $\langle \widetilde Q \, A_e \rangle$. Repeatedly averaging 9 different, random positions (with replacement) results in about 20\% variation on their averaged $S_{12}$ coefficients at each frequency, allowing us to conclude 9 positions and polarizations  provides acceptable convergence. 


\subsection{Correction and uncertainty of \texorpdfstring{$\langle \widetilde{Q} \, A_e \rangle$}{<QAe>}}
\label{sec:simulationUncertainty}
As outlined above, we approximate the uncertainty of the simulation by injecting power into the system via a second antenna and comparing the results to simulation. 

For a two port microwave device, the ratio between the voltage presented at port one and the voltage measured at port two is known as $S_{21}$. For our system, $S_{21}$ is a measurable quantity which is similar to a dark photon detection in that it requires the antenna to convert an electric field (which has interacted with the room) into a port voltage. Having frequency dependent measurements of $S_{21}$ for simulation and measurement give us a correction to the simulation (to account for discrepancies in geometry) and estimate the uncertainty on $\langle \widetilde{Q} \, A_e \rangle$. 

\noindent The difference between the measured and simulated values of $\langle \lvert S_{21} \rvert \rangle$ can be described by

\begin{equation}
\label{eq:alphaCorrection}
\langle \lvert S_{21}^{\textrm{meas}} \rvert ^{2} \rangle = \alpha \langle \lvert S_{21}^{\textrm{sim}} \rvert ^{2} \rangle,
\end{equation}

\noindent where meas/sim indicates measured/simulated and the average is over all 18 measured/simulated positions and orientations of the antenna. We have taken the square since we are interested in the aperture, which is proportional to the square of the voltage. This equation implies $\alpha$ is a frequency dependent, multiplicative correction factor which results in a corrected $\langle \lvert S_{21}^{\textrm{sim}} \rvert ^{2} \rangle$. We find $\alpha$ to have a mean of 0.6, a minimum of 0.1 and a maximum of 2.

To determine uncertainty on effective aperture, we define the following test statistic
\begin{equation}
\Delta = \frac{\langle \lvert S^{\textrm{meas}}_{21, n} \rvert ^{2} \rangle - \alpha \langle \lvert S^{\textrm{sim}}_{21, n} \rvert ^{2} \rangle }{\langle \lvert S^{\textrm{meas}}_{21} \rvert ^{2} \rangle }, 
\end{equation}
\noindent where $n$ refers to the subset of $n$ measured/simulated positions sampled randomly with replacement. $\Delta$ defines the fractional difference between corrected, simulated $S_{21}$ and measured $S_{21}$. The test statistic, $\Delta$, is calculated 1000 times, providing a distribution of frequency dependent $\Delta$s. The curves bounding 63\% of these curves are taken to be the uncertainty on $\Delta$.

To convert back to uncertainty on antenna aperture, we note that the fractional uncertainty is related back to the match parameter as

\begin{equation}
\delta \langle \widetilde{Q} \, A_e \rangle =  \langle \widetilde{Q} \, A_e \rangle  \, \delta \Delta .
\end{equation}

\begin{figure}[H]
\includegraphics[width=0.48\textwidth]{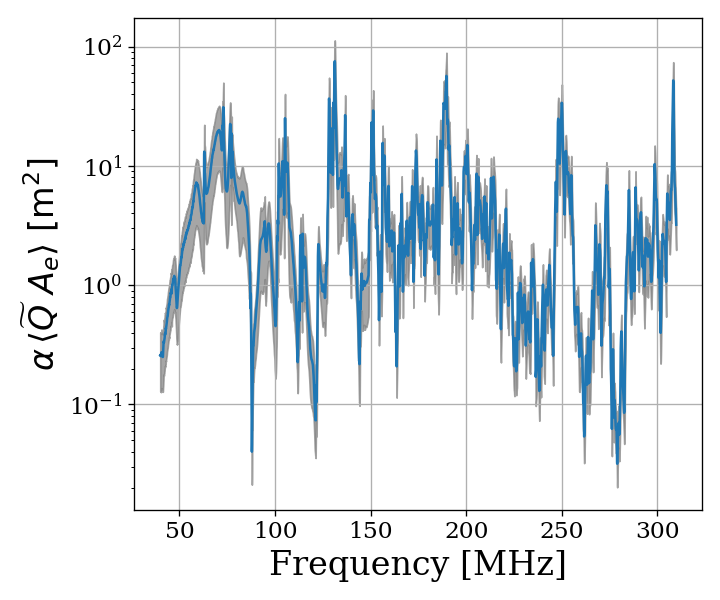}
  \caption{Corrected average effective aperture. Calculated with COMSOL RF. The aperture correction $\alpha$ (Eq. \ref{eq:alphaCorrection}) and its uncertainty (\textit{gray}) are estimated by comparing simulations to measured $S$ parameters.}
  \label{fig:aperature}
\end{figure}

This corrected average effective aperture as a function of frequency is shown in Fig. \ref{fig:aperature}.

A brief summary of the system's aperture is in order. In free space an antenna's ability to couple an incoming wave's power density into a transmission line is given by it's effective aperture, Eq. \ref{eq:effAppFreeSpace}. An antenna in a cavity acts as a coupled oscillator which exhibits very complex resonances above the first few modes (around 100\,MHz for our system). Attempts to simulate an aperture for the antenna-cavity system are difficult because of the system's extreme dependence on placement of any conductor in the room, especially the antenna. Averaging over system configurations (antenna positions and polarizations in our case) allows for a significantly more repeatable \emph{statistical} treatment of the aperture/quality factor, which we call $\langle \widetilde{Q} \, A_e \rangle$. Comparison of simulated and measured $S_{21}$ gives a small, dimensionless correction factor $\alpha$, Eq. \ref{eq:alphaCorrection}. 

Armed with $\alpha\,\langle \widetilde{Q} \, A_e \rangle$ we are now able to compute a limit on epsilon using measured and simulated quantities via Eqs. \ref{eq:pLimAnt} and \ref{eq:powerVsEobs},


\begin{equation}
\label{eq:epsilonLim}
    \epsilon(\nu) <
   \sqrt{ \frac{1}{2\,c \,\rho_\mathrm{DM}} \,  \, \frac{\mathrm{P}_\textrm{ant}^{\mathrm{lim}}}{\alpha \,\langle \widetilde Q A_{e}\rangle }} ,
\end{equation}

\noindent where $c$ is the speed of light, $\rho_\mathrm{DM}$ is the local dark matter density and $\mathrm{P}_\mathrm{ant}^\mathrm{lim}$ is defined in Eq. \ref{eq:pLimAnt}. We have separated the equation into constants (or in the case of $\rhoDM$, values which we fix) and values which we measure or simulate.

In order to validate our entire detection system, we inject sub-threshold signals into the shielded room to verify we are able to detect them.

\subsection{Hardware RF injection tests} 
Several hardware injection tests were performed at various frequencies using a quarter wavelength monopole inside the shielded room. Each injection was a sub-threshold monochromatic signal buried deep under the noise. The standard deviation of the noise was shown to average down with the square root of integration time, revealing the injected signal using the detection algorithm described in Sec. \ref{sec:analysis}. These hardware tests are simple injections of pure tones which show that the entire experiment (antenna through data analysis) is able to detect small electric fields present in the cavity after averaging the noise for a predictable amount of time. An interesting method for injecting realistic line shapes is discussed by Zhu et. al \cite{Zhu_2023}. Such simple hardware injection tests should not be confused with the Monte Carlo's software injection of realistic dark photon line shapes onto a background of known standard deviation. 

\section{Results}
\label{sec:results}

\begin{figure}
\includegraphics[width=0.48\textwidth]{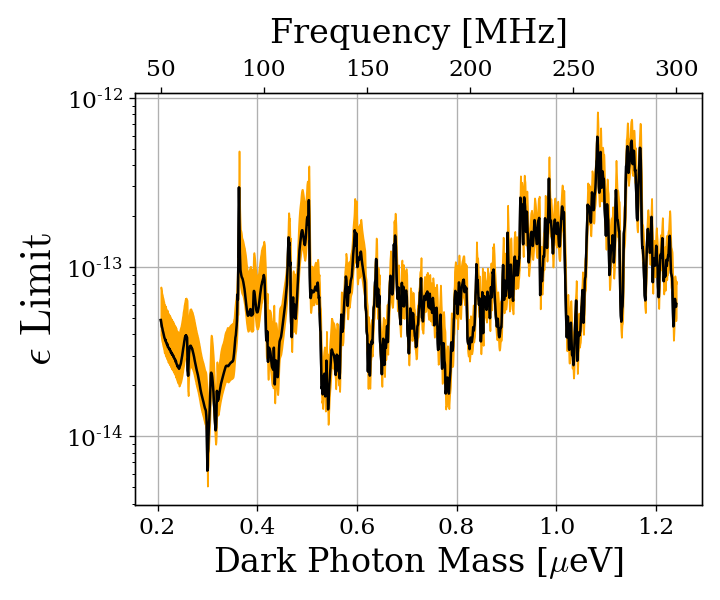}
  \caption{95\,\% exclusion limit on $\epsilon$ with uncertainty shown in orange shaded region. This is based on a local dark matter density of $\rhoDM = 0.45\,\mathrm{GeV/cm^3}$. The error estimate does not take the comparatively small gain and amplifier noise temperature errors into account.}
  \label{fig:EpsilonLimitPlotZoom}
\end{figure}

\begin{figure*}
\includegraphics[width=1.0\textwidth]{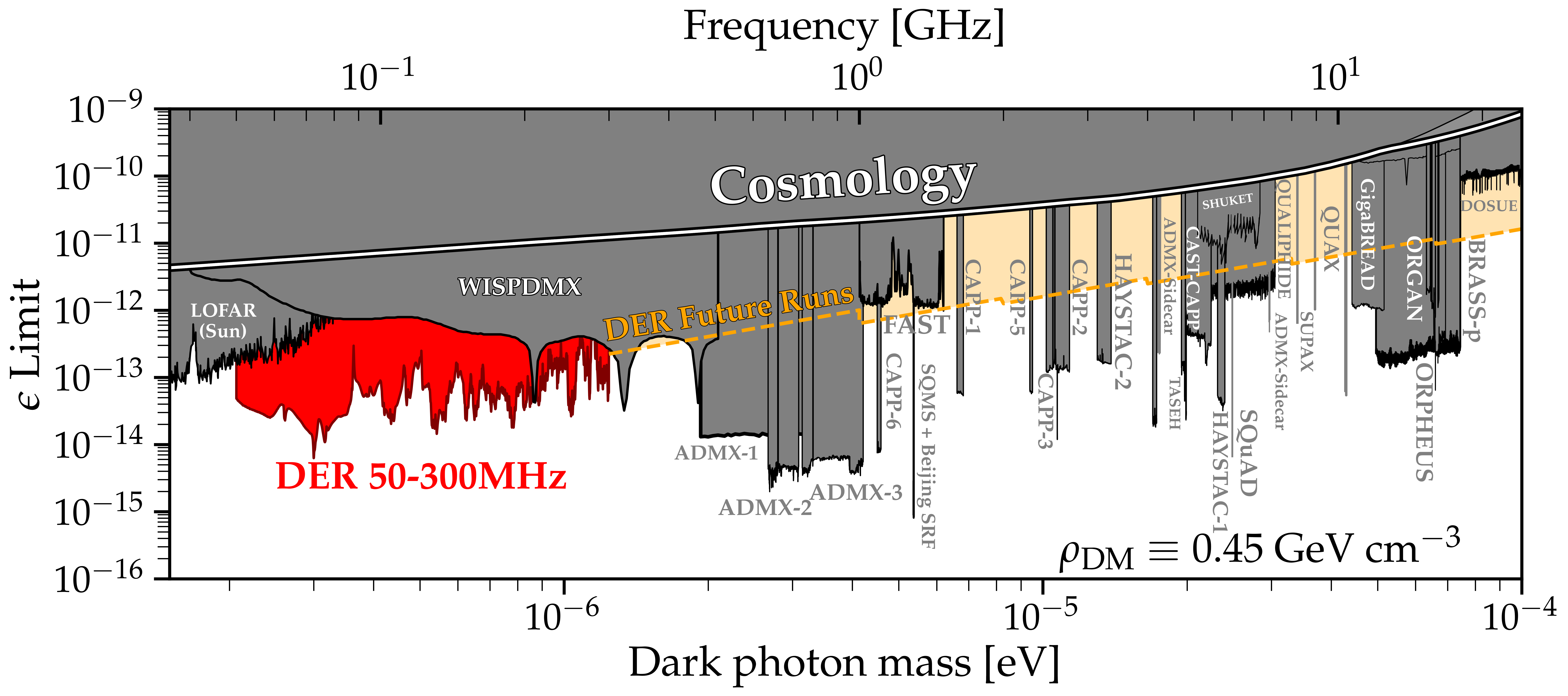}
\caption{Dark photon limits of various experiments with this work shown in red. The ragged lower bound is due to the complex structure of the resonant modes of the shielded room. Plot adapted from \cite{DarkPhotonHandbook} and includes limit projections of various axion experiments. Astrophysical limits such as CMB interactions with the dark photon are in the region labeled \textit{Cosmology}. Planned wideband extensions of our experiment search from 0.3-25\,GHz in similar room temperature RF enclosures are indicated (\textit{yellow}).}
\label{fig:EpsilonLimitPlot}
\end{figure*}

In this section, we report a 95\%, frequency-dependent, exclusion limit on the kinetic mixing strength $\epsilon$ of the dark photon (Fig.~\ref{fig:EpsilonLimitPlotZoom}).  We discuss uncertainties on measured data, identification of a candidate signal and our process to exclude it.  Finally, we display our results in context by plotting these new limits on top of an aggregation of existing limits in Fig.~\ref{fig:EpsilonLimitPlot}. Future runs of this experiment from 0.3-25\,GHz in similar room temperature RF enclosures and 100\,K noise temperature LNAs are indicated. We have only indicated planned runs, however at microwave frequencies, highly resonant cryogenic cavities and cryogenic LNAs as well as sub-THz instrumentation are feasible and could result in an order of magnitude improvement in the limit over the indicated frequency range and beyond.

\subsection{Discussion of uncertainties}
\label{sec:ResultsUncertaincities}

The systematic uncertainty in this experiment comes primarily from three sources, listed in order of their contribution from greatest to least:
\begin{enumerate}
    \item Fractional uncertainty on the simulated antenna aperture, which is discussed in \ref{sec:simulationUncertainty}, $\approx$ 60\,\%
    \item Fractional uncertainty on the first-stage amplifier noise temperature, $\approx$ 10\,\%
    \item Fractional uncertainty on the gain of the amplifier chain, $\approx$ 5\,\%
\end{enumerate}

The uncertainty on the simulated antenna aperture is significantly larger than the other two, and so we neglect them in the uncertainty in the $\epsilon$ limit. 

We follow the convention of similar experiments where we fix the value of $\rhoDM$ and solve for an $\epsilon$ limit given this value. Therefore we treat $\rhoDM$ as a known constant with no uncertainty. 

\subsection{Rejection of a single candidate}
\label{sec:falsePosSig}
Passing $\mathrm{S}_\textbf{o}$ through the detection algorithm diagrammed in Fig.~\ref{fig:detectorFlowChart} yields a single candidate at 299.97\,MHz which is approximately 1\,kHz wide. This candidate first became detectable above the noise after about 4 days of averaging, indicating it is just on the threshold of what we are able to detect. Four factors cause us to conclude the candidate is an interfering signal originating from within the PC or ADC, allowing us to remove it:

\begin{itemize}
    \item The candidate is present not only in the main spectrum but also the veto and terminator spectrum
    \item Inspection of the time evolution of this signal shows a narrow signal (about two bins, or 100\,Hz wide) which seems to wander in frequency periodically over the course of a day and therefore with temperature. This is expected behavior for a quartz oscillator
    \item  Reducing the gain of the system causes the SNR of the candidate to \textit{increase}, indicating it enters the signal path after the gain stages
    \item Changing the clock rate causes the frequency of the candidate to change
\end{itemize}

\newpage
\section{Discussion}
\label{sec:discussion}

This experiment extends the earlier results of our pilot experiment \cite{GroupPaper}, which was designed to demonstrate feasibility of the Dark E-field Radio technique. The pilot experiment was run over the same frequency range as the experiment reported here, but did not make use of the calibration techniques to approximate statistical uniformity, nor did it fully account for the resonant enhancement of the cavity. In this paper we describe how we randomize antenna positions by moving it many times during the run. In addition, we detail EM simulations which give the average relation between the E-field at the antenna and the voltage into the LNA, accounting for resonant enhancement of the cavity. A $2^{24}$-point FFT produces a spectrum dominated by background thermal noise which varies gradually with frequency.

We then searched over the full 50-300\,MHz frequency span for any narrow-band dark photon signal of at least 5\% global significance. Optimally filtering the resulting spectrum, we detect a single candidate which we are able to identify as interference, likely from our electronics. Rejecting this candidate, we obtain a null result for any signal which could be attributed to the dark photon in our frequency range. The resulting 95\% exclusion limit for the dark photon kinetic coupling $\epsilon$ is then obtained over this mass range of 0.2-1.2\,$\mu$eV. Our null result is a factor of $\approx$\,100 more sensitive than current astrophysical limits.

Ultimately, we can apply this detection technique at higher frequencies, ultimately going up to the sub-THz band. This will require new antennas and microwave electronics. Cryogenic cavities and LNAs could improve our sensitivity by an order of magnitude.


\section*{Acknowledgments}
\label{sec:Acknoledgments}

We acknowledge helpful discussions with John Conway, Matthew Citron, Gabriel Lynch, Markus Luty, Umut {\"O}ktem, and Greg Wright. We gratefully acknowledge the early COMSOL simulation work of Vasco Rodriguez. This work was partially supported by the Brinson Foundation and DOE grant DE-SC0009999. BG was supported by the NSSC program funded by DOE/NNSA award number DE-NA0000979. We also received partial support from the HEPCAT program under DOE award number DE-SC0022313.

\begin{table*}[ht]
\begin{center}
\begin{tabular}{cccc}
\hline
Symbol & Units & Meaning & Section Introduced \\
\hline
$A_e(\nu)$ &$m^2$ & Effective aperture & \ref{sec:avgEffAperture}
\\
$B(\nu)$ & W/Hz & Background shape of $\mathrm{S}_\textbf{o}$&  \ref{sec:FitB}\\
$\hat{B}(\nu)$ & W/Hz & Fit to $\mathrm{S}_\textbf{o}$. Estimate of B &  \ref{sec:FitB} \\
$\mathbf{E}_{\mathrm{ant}}$ & V/m & Observable electric field at antenna in free space & \ref{sec:intro} 
\\
N & None & Number of bins&  \ref{sec:normalizedSpectrum}
\\
$\mathrm{P}_\textrm{ant}^{\mathrm{lim}}(\nu)$ & W & 95\,\% exclusion limit on power due to dark photon (antenna referred)&  \ref{sec:calibration}\\
$\mathrm{S}_{\mathrm{ant}}(\nu)$ & W/Hz & Antenna Noise &   \ref{sec:BackgroundNoiseOverview}
\\
$\mathrm{S}_\textbf{i}(\nu)$ & W/Hz &Input-referred power spectral density &  \ref{sec:BackgroundNoiseOverview} \\
$\mathrm{S}_\textbf{o}(\nu)$ & W/Hz & Output-referred power spectral density &  \ref{sec:BackgroundNoiseOverview}
 \\
 SE$(\nu)$ & dB & Shielding effectiveness & \ref{sec:BackgroundNoiseOverview}
\\
$Q_\mathrm{DP}$ & None & Quality factor of DP&  \ref{sec:freqResolution}
\\
$\widetilde{Q}(\nu)$ & None & Composite quality factor of antenna/room system&  \ref{sec:avgEffAperture}
\\
$z$ & None & Number of STD&  \ref{sec:normalizedSpectrum}
\\
$\alpha(\nu)$ & None & Correction factor for simulation & \ref{sec:simulationUncertainty}
\\
$\Delta(\nu)$ & None & Test statistic to estimate simulation uncertainty &  \ref{sec:simulationUncertainty}
\\
$\Delta \nu_B$ & Hz& Width feature on background&  \ref{sec:FitB}
\\
$\DeltanuRF$ & Hz& Width of bin&  \ref{sec:freqResolution}
\\
$\Delta \nu_\mathrm{DP}(\nu)$ & Hz& Width of dark photon signal&  \ref{sec:FitB}
\\
$\epsilon(\nu)$ & None & Kinetic mixing parameter &   \ref{sec:intro}
\\
$\mu_\mathrm{norm}$ & None & Normalized mean&  \ref{sec:normalizedSpectrum}
\\
$\sigma_\mathrm{norm}$ & None & Normalized STD&  \ref{sec:normalizedSpectrum}
\\
$\tau$ & s & Total integration time&  \ref{sec:experimentalRequirements}
\\
\hline

\end{tabular}
\end{center}
\caption{Parameter definitions used throughout the paper. Frequency dependence denoted by $(\nu)$. We remove this explicit dependence after the first time it appears throughout the paper. }
\label{table:spectrum_labels}
\end{table*}

\clearpage
\bibliography{darkradio}

\providecommand{\noopsort}[1]{}\providecommand{\singleletter}[1]{#1}%
\begin{thebibliography}{34}%
\makeatletter
\providecommand \@ifxundefined [1]{%
 \@ifx{#1\undefined}
}%
\providecommand \@ifnum [1]{%
 \ifnum #1\expandafter \@firstoftwo
 \else \expandafter \@secondoftwo
 \fi
}%
\providecommand \@ifx [1]{%
 \ifx #1\expandafter \@firstoftwo
 \else \expandafter \@secondoftwo
 \fi
}%
\providecommand \natexlab [1]{#1}%
\providecommand \enquote  [1]{``#1''}%
\providecommand \bibnamefont  [1]{#1}%
\providecommand \bibfnamefont [1]{#1}%
\providecommand \citenamefont [1]{#1}%
\providecommand \href@noop [0]{\@secondoftwo}%
\providecommand \href [0]{\begingroup \@sanitize@url \@href}%
\providecommand \@href[1]{\@@startlink{#1}\@@href}%
\providecommand \@@href[1]{\endgroup#1\@@endlink}%
\providecommand \@sanitize@url [0]{\catcode `\\12\catcode `\$12\catcode
  `\&12\catcode `\#12\catcode `\^12\catcode `\_12\catcode `\%12\relax}%
\providecommand \@@startlink[1]{}%
\providecommand \@@endlink[0]{}%
\providecommand \url  [0]{\begingroup\@sanitize@url \@url }%
\providecommand \@url [1]{\endgroup\@href {#1}{\urlprefix }}%
\providecommand \urlprefix  [0]{URL }%
\providecommand \Eprint [0]{\href }%
\providecommand \doibase [0]{https://doi.org/}%
\providecommand \selectlanguage [0]{\@gobble}%
\providecommand \bibinfo  [0]{\@secondoftwo}%
\providecommand \bibfield  [0]{\@secondoftwo}%
\providecommand \translation [1]{[#1]}%
\providecommand \BibitemOpen [0]{}%
\providecommand \bibitemStop [0]{}%
\providecommand \bibitemNoStop [0]{.\EOS\space}%
\providecommand \EOS [0]{\spacefactor3000\relax}%
\providecommand \BibitemShut  [1]{\csname bibitem#1\endcsname}%
\let\auto@bib@innerbib\@empty
\bibitem [{\citenamefont {Misiaszek}\ and\ \citenamefont
  {Rossi}(2024)}]{misiaszek2024direct}%
  \BibitemOpen
  \bibfield  {author} {\bibinfo {author} {\bibfnamefont {M.}~\bibnamefont
  {Misiaszek}}\ and\ \bibinfo {author} {\bibfnamefont {N.}~\bibnamefont
  {Rossi}},\ }\href {https://doi.org/10.3390/sym16020201} {\bibfield  {journal}
  {\bibinfo  {journal} {Symmetry}\ }\textbf {\bibinfo {volume} {16}},\ \bibinfo
  {pages} {201} (\bibinfo {year} {2024})}\BibitemShut {NoStop}%
\bibitem [{\citenamefont {Battaglieri}\ \emph {et~al.}(2017)\citenamefont
  {Battaglieri} \emph {et~al.}}]{Battaglieri:2017aum}%
  \BibitemOpen
  \bibfield  {author} {\bibinfo {author} {\bibfnamefont {M.}~\bibnamefont
  {Battaglieri}} \emph {et~al.},\ }in\ \href@noop {} {\emph {\bibinfo
  {booktitle} {{U.S. Cosmic Visions: New Ideas in Dark Matter}}}}\ (\bibinfo
  {year} {2017})\ \Eprint {https://arxiv.org/abs/1707.04591} {arXiv:1707.04591
  [hep-ph]} \BibitemShut {NoStop}%
\bibitem [{\citenamefont {Butler}\ \emph {et~al.}(2023)\citenamefont {Butler},
  \citenamefont {Chivukula}, \citenamefont {de~Gouvêa}, \citenamefont {Han},
  \citenamefont {Kim}, \citenamefont {Cushman}, \citenamefont {Farrar},
  \citenamefont {Kolomensky}, \citenamefont {Nagaitsev}, \citenamefont {Yunes}
  \emph {et~al.}}]{butler2023report}%
  \BibitemOpen
  \bibfield  {author} {\bibinfo {author} {\bibfnamefont {J.~N.}\ \bibnamefont
  {Butler}}, \bibinfo {author} {\bibfnamefont {R.~S.}\ \bibnamefont
  {Chivukula}}, \bibinfo {author} {\bibfnamefont {A.}~\bibnamefont
  {de~Gouvêa}}, \bibinfo {author} {\bibfnamefont {T.}~\bibnamefont {Han}},
  \bibinfo {author} {\bibfnamefont {Y.-K.}\ \bibnamefont {Kim}}, \bibinfo
  {author} {\bibfnamefont {P.}~\bibnamefont {Cushman}}, \bibinfo {author}
  {\bibfnamefont {G.~R.}\ \bibnamefont {Farrar}}, \bibinfo {author}
  {\bibfnamefont {Y.~G.}\ \bibnamefont {Kolomensky}}, \bibinfo {author}
  {\bibfnamefont {S.}~\bibnamefont {Nagaitsev}}, \bibinfo {author}
  {\bibfnamefont {N.}~\bibnamefont {Yunes}}, \emph {et~al.},\ }\href@noop {}
  {\bibinfo {title} {Report of the 2021 u.s. community study on the future of
  particle physics (snowmass 2021) summary chapter}} (\bibinfo {year} {2023}),\
  \Eprint {https://arxiv.org/abs/2301.06581} {arXiv:2301.06581 [hep-ex]}
  \BibitemShut {NoStop}%
\bibitem [{\citenamefont {Nelson}\ and\ \citenamefont
  {Scholtz}(2011)}]{PhysRevD.84.103501}%
  \BibitemOpen
  \bibfield  {author} {\bibinfo {author} {\bibfnamefont {A.~E.}\ \bibnamefont
  {Nelson}}\ and\ \bibinfo {author} {\bibfnamefont {J.}~\bibnamefont
  {Scholtz}},\ }\href {https://doi.org/10.1103/PhysRevD.84.103501} {\bibfield
  {journal} {\bibinfo  {journal} {Phys. Rev. D}\ }\textbf {\bibinfo {volume}
  {84}},\ \bibinfo {pages} {103501} (\bibinfo {year} {2011})}\BibitemShut
  {NoStop}%
\bibitem [{\citenamefont {Graham}\ \emph {et~al.}(2016)\citenamefont {Graham},
  \citenamefont {Mardon},\ and\ \citenamefont {Rajendran}}]{Graham:2015rva}%
  \BibitemOpen
  \bibfield  {author} {\bibinfo {author} {\bibfnamefont {P.}~\bibnamefont
  {Graham}}, \bibinfo {author} {\bibfnamefont {J.}~\bibnamefont {Mardon}},\
  and\ \bibinfo {author} {\bibfnamefont {S.}~\bibnamefont {Rajendran}},\ }\href
  {https://arxiv.org/abs/1504.02102} {\bibfield  {journal} {\bibinfo  {journal}
  {Phys. Rev. D}\ }\textbf {\bibinfo {volume} {93}},\ \bibinfo {pages} {103520}
  (\bibinfo {year} {2016})}\BibitemShut {NoStop}%
\bibitem [{\citenamefont {{Holdom}}(1986)}]{1986PhLB..166..196H}%
  \BibitemOpen
  \bibfield  {author} {\bibinfo {author} {\bibfnamefont {B.}~\bibnamefont
  {{Holdom}}},\ }\href {https://doi.org/10.1016/0370-2693(86)91377-8}
  {\bibfield  {journal} {\bibinfo  {journal} {Phys. Lett. B}\ }\textbf
  {\bibinfo {volume} {166}},\ \bibinfo {pages} {196} (\bibinfo {year}
  {1986})}\BibitemShut {NoStop}%
\bibitem [{\citenamefont {Read}(2014)}]{Read:2014qva}%
  \BibitemOpen
  \bibfield  {author} {\bibinfo {author} {\bibfnamefont {J.}~\bibnamefont
  {Read}},\ }\href {https://arxiv.org/abs/1404.1938} {\bibfield  {journal}
  {\bibinfo  {journal} {J. Phys.}\ }\textbf {\bibinfo {volume} {G41}},\
  \bibinfo {pages} {063101} (\bibinfo {year} {2014})}\BibitemShut {NoStop}%
\bibitem [{\citenamefont {{de Salas}}(2020)}]{2019arXiv191014366D}%
  \BibitemOpen
  \bibfield  {author} {\bibinfo {author} {\bibfnamefont {P.}~\bibnamefont {{de
  Salas}}},\ }\href {https://arxiv.org/abs/1910.14366} {\bibfield  {journal}
  {\bibinfo  {journal} {J. Phys. Conf. Ser.}\ }\textbf {\bibinfo {volume}
  {1468}},\ \bibinfo {pages} {012020} (\bibinfo {year} {2020})}\BibitemShut
  {NoStop}%
\bibitem [{\citenamefont {de~Salas}\ and\ \citenamefont
  {Widmark}(2021)}]{de2021dark}%
  \BibitemOpen
  \bibfield  {author} {\bibinfo {author} {\bibfnamefont {P.~F.}\ \bibnamefont
  {de~Salas}}\ and\ \bibinfo {author} {\bibfnamefont {A.}~\bibnamefont
  {Widmark}},\ }\href {https://doi.org/10.1088/1361-6633/ac24e7} {\bibfield
  {journal} {\bibinfo  {journal} {Reports on Progress in Physics}\ }\textbf
  {\bibinfo {volume} {84}},\ \bibinfo {pages} {104901} (\bibinfo {year}
  {2021})}\BibitemShut {NoStop}%
\bibitem [{\citenamefont {Chaudhuri}\ \emph {et~al.}(2015)\citenamefont
  {Chaudhuri}, \citenamefont {Graham}, \citenamefont {Irwin}, \citenamefont
  {Mardon}, \citenamefont {Rajendran},\ and\ \citenamefont
  {Zhao}}]{Chaudhuri_2015}%
  \BibitemOpen
  \bibfield  {author} {\bibinfo {author} {\bibfnamefont {S.}~\bibnamefont
  {Chaudhuri}}, \bibinfo {author} {\bibfnamefont {P.}~\bibnamefont {Graham}},
  \bibinfo {author} {\bibfnamefont {K.}~\bibnamefont {Irwin}}, \bibinfo
  {author} {\bibfnamefont {J.}~\bibnamefont {Mardon}}, \bibinfo {author}
  {\bibfnamefont {S.}~\bibnamefont {Rajendran}},\ and\ \bibinfo {author}
  {\bibfnamefont {Y.}~\bibnamefont {Zhao}},\ }\href
  {https://arxiv.org/abs/1411.7382} {\bibfield  {journal} {\bibinfo  {journal}
  {Phys. Rev. D}\ }\textbf {\bibinfo {volume} {92}},\ \bibinfo {pages} {075012}
  (\bibinfo {year} {2015})}\BibitemShut {NoStop}%
\bibitem [{\citenamefont {Gramolin}\ \emph {et~al.}(2022)\citenamefont
  {Gramolin}, \citenamefont {Wickenbrock}, \citenamefont {Aybas}, \citenamefont
  {Bekker}, \citenamefont {Budker}, \citenamefont {Centers}, \citenamefont
  {Figueroa}, \citenamefont {Kimball},\ and\ \citenamefont
  {Sushkov}}]{Gramolin_2022}%
  \BibitemOpen
  \bibfield  {author} {\bibinfo {author} {\bibfnamefont {A.~V.}\ \bibnamefont
  {Gramolin}}, \bibinfo {author} {\bibfnamefont {A.}~\bibnamefont
  {Wickenbrock}}, \bibinfo {author} {\bibfnamefont {D.}~\bibnamefont {Aybas}},
  \bibinfo {author} {\bibfnamefont {H.}~\bibnamefont {Bekker}}, \bibinfo
  {author} {\bibfnamefont {D.}~\bibnamefont {Budker}}, \bibinfo {author}
  {\bibfnamefont {G.~P.}\ \bibnamefont {Centers}}, \bibinfo {author}
  {\bibfnamefont {N.~L.}\ \bibnamefont {Figueroa}}, \bibinfo {author}
  {\bibfnamefont {D.~F.~J.}\ \bibnamefont {Kimball}},\ and\ \bibinfo {author}
  {\bibfnamefont {A.~O.}\ \bibnamefont {Sushkov}},\ }\bibfield  {journal}
  {\bibinfo  {journal} {Physical Review D}\ }\textbf {\bibinfo {volume}
  {105}},\ \href {https://doi.org/10.1103/physrevd.105.035029}
  {10.1103/physrevd.105.035029} (\bibinfo {year} {2022})\BibitemShut {NoStop}%
\bibitem [{AB-(2019)}]{AB-900ADatasheet}%
  \BibitemOpen
  \href {https://documentation.com-power.com/pdf/AB-900A-1.pdf} {\emph
  {\bibinfo {title} {Biconical Antenna AB-900A}}},\ \bibinfo {organization}
  {Com-Power Corporation} (\bibinfo {year} {2019}),\ \bibinfo {note} {rev.
  D05.19}\BibitemShut {NoStop}%
\bibitem [{Lin(2021)}]{LindgrenDatasheet}%
  \BibitemOpen
  \href
  {https://www.ets-lindgren.com/datasheet/shielding/rf-shielding-and-accessories/11003/1100307}
  {\emph {\bibinfo {title} {Series 83 RF Shielded Enclosure}}},\ \bibinfo
  {organization} {Lindgren R.F. Enclosures, Inc.} (\bibinfo {year}
  {2021})\BibitemShut {NoStop}%
\bibitem [{\citenamefont {Godfrey}\ \emph {et~al.}(2021)\citenamefont
  {Godfrey}, \citenamefont {Tyson}, \citenamefont {Hillbrand}, \citenamefont
  {Balajthy}, \citenamefont {Polin}, \citenamefont {Tripathi}, \citenamefont
  {Klomp}, \citenamefont {Levine}, \citenamefont {MacFadden}, \citenamefont
  {Kolner} \emph {et~al.}}]{GroupPaper}%
  \BibitemOpen
  \bibfield  {author} {\bibinfo {author} {\bibfnamefont {B.}~\bibnamefont
  {Godfrey}}, \bibinfo {author} {\bibfnamefont {J.~A.}\ \bibnamefont {Tyson}},
  \bibinfo {author} {\bibfnamefont {S.}~\bibnamefont {Hillbrand}}, \bibinfo
  {author} {\bibfnamefont {J.}~\bibnamefont {Balajthy}}, \bibinfo {author}
  {\bibfnamefont {D.}~\bibnamefont {Polin}}, \bibinfo {author} {\bibfnamefont
  {S.~M.}\ \bibnamefont {Tripathi}}, \bibinfo {author} {\bibfnamefont
  {S.}~\bibnamefont {Klomp}}, \bibinfo {author} {\bibfnamefont
  {J.}~\bibnamefont {Levine}}, \bibinfo {author} {\bibfnamefont
  {N.}~\bibnamefont {MacFadden}}, \bibinfo {author} {\bibfnamefont {B.~H.}\
  \bibnamefont {Kolner}}, \emph {et~al.},\ }\href
  {https://doi.org/10.1103/PhysRevD.104.012013} {\bibfield  {journal} {\bibinfo
   {journal} {Phys. Rev. D}\ }\textbf {\bibinfo {volume} {104}},\ \bibinfo
  {pages} {012013} (\bibinfo {year} {2021})}\BibitemShut {NoStop}%
\bibitem [{\citenamefont {Hill}(1998{\natexlab{a}})}]{NBSTechNote1506}%
  \BibitemOpen
  \bibfield  {author} {\bibinfo {author} {\bibfnamefont {D.~A.}\ \bibnamefont
  {Hill}},\ }\href@noop {} {\emph {\bibinfo {title} {Electromagnetic theory of
  reverberation chambers}}},\ \bibinfo {type} {Tech. Rep.}\ (\bibinfo
  {institution} {United States. Government Printing Office},\ \bibinfo {year}
  {1998})\BibitemShut {NoStop}%
\bibitem [{\citenamefont {Ladbury}\ \emph {et~al.}(1999)\citenamefont
  {Ladbury}, \citenamefont {Koepke},\ and\ \citenamefont
  {Camell}}]{NBSTechNote1508}%
  \BibitemOpen
  \bibfield  {author} {\bibinfo {author} {\bibfnamefont {J.}~\bibnamefont
  {Ladbury}}, \bibinfo {author} {\bibfnamefont {G.}~\bibnamefont {Koepke}},\
  and\ \bibinfo {author} {\bibfnamefont {D.}~\bibnamefont {Camell}},\
  }\href@noop {} {\bibinfo {title} {Evaluation of the nasa langley research
  center mode-stirred chamber facility}} (\bibinfo {year} {1999})\BibitemShut
  {NoStop}%
\bibitem [{\citenamefont {Crawford}\ and\ \citenamefont
  {Koepke}(1986)}]{NBSTechnote1092}%
  \BibitemOpen
  \bibfield  {author} {\bibinfo {author} {\bibfnamefont {M.~L.}\ \bibnamefont
  {Crawford}}\ and\ \bibinfo {author} {\bibfnamefont {G.~H.}\ \bibnamefont
  {Koepke}},\ }\href@noop {} {\emph {\bibinfo {title} {Design, evaluation, and
  use of a reverberation chamber for performing electromagnetic
  susceptibility/vulnerability measurements}}},\ \bibinfo {type} {Tech. Rep.}\
  (\bibinfo  {institution} {United States. Government Printing Office.},\
  \bibinfo {year} {1986})\BibitemShut {NoStop}%
\bibitem [{\citenamefont {Tai}(1961)}]{TaiEffectiveApertureOriginal}%
  \BibitemOpen
  \bibfield  {author} {\bibinfo {author} {\bibfnamefont {C.}~\bibnamefont
  {Tai}},\ }\href {https://doi.org/10.1109/TAP.1961.1144972} {\bibfield
  {journal} {\bibinfo  {journal} {IRE Transactions on Antennas and
  Propagation}\ }\textbf {\bibinfo {volume} {9}},\ \bibinfo {pages} {224}
  (\bibinfo {year} {1961})}\BibitemShut {NoStop}%
\bibitem [{\citenamefont {Serra}\ \emph {et~al.}(2017)\citenamefont {Serra},
  \citenamefont {Marvin}, \citenamefont {Moglie}, \citenamefont {Primiani},
  \citenamefont {Cozza}, \citenamefont {Arnaut}, \citenamefont {Huang},
  \citenamefont {Hatfield}, \citenamefont {Klingler},\ and\ \citenamefont
  {Leferink}}]{ReverbChambersALaCarte}%
  \BibitemOpen
  \bibfield  {author} {\bibinfo {author} {\bibfnamefont {R.}~\bibnamefont
  {Serra}}, \bibinfo {author} {\bibfnamefont {A.~C.}\ \bibnamefont {Marvin}},
  \bibinfo {author} {\bibfnamefont {F.}~\bibnamefont {Moglie}}, \bibinfo
  {author} {\bibfnamefont {V.~M.}\ \bibnamefont {Primiani}}, \bibinfo {author}
  {\bibfnamefont {A.}~\bibnamefont {Cozza}}, \bibinfo {author} {\bibfnamefont
  {L.~R.}\ \bibnamefont {Arnaut}}, \bibinfo {author} {\bibfnamefont
  {Y.}~\bibnamefont {Huang}}, \bibinfo {author} {\bibfnamefont {M.~O.}\
  \bibnamefont {Hatfield}}, \bibinfo {author} {\bibfnamefont {M.}~\bibnamefont
  {Klingler}},\ and\ \bibinfo {author} {\bibfnamefont {F.}~\bibnamefont
  {Leferink}},\ }\href {https://doi.org/10.1109/MEMC.2017.7931986} {\bibfield
  {journal} {\bibinfo  {journal} {IEEE Electromagnetic Compatibility Magazine}\
  }\textbf {\bibinfo {volume} {6}},\ \bibinfo {pages} {63} (\bibinfo {year}
  {2017})}\BibitemShut {NoStop}%
\bibitem [{\citenamefont {Hill}(2009)}]{hill2009}%
  \BibitemOpen
  \bibfield  {author} {\bibinfo {author} {\bibfnamefont {D.}~\bibnamefont
  {Hill}},\ }\href {https://doi.org/10.1002/9780470495056} {\emph {\bibinfo
  {title} {Electromagnetic Fields in Cavities}}}\ (\bibinfo  {publisher}
  {Wiley},\ \bibinfo {address} {Hoboken, N.J.},\ \bibinfo {year}
  {2009})\BibitemShut {NoStop}%
\bibitem [{\citenamefont {Hill}(1998{\natexlab{b}})}]{hillPlaneWaves}%
  \BibitemOpen
  \bibfield  {author} {\bibinfo {author} {\bibfnamefont {D.}~\bibnamefont
  {Hill}},\ }\href {https://doi.org/10.1109/15.709418} {\bibfield  {journal}
  {\bibinfo  {journal} {IEEE Transactions on Electromagnetic Compatibility}\
  }\textbf {\bibinfo {volume} {40}},\ \bibinfo {pages} {209} (\bibinfo {year}
  {1998}{\natexlab{b}})}\BibitemShut {NoStop}%
\bibitem [{\citenamefont {Dicke}(1946)}]{Dicke:1946}%
  \BibitemOpen
  \bibfield  {author} {\bibinfo {author} {\bibfnamefont {R.~H.}\ \bibnamefont
  {Dicke}},\ }\href {https://doi.org/10.1063/1.1770483} {\bibfield  {journal}
  {\bibinfo  {journal} {Rev. Sci. Instrum.}\ }\textbf {\bibinfo {volume}
  {17}},\ \bibinfo {pages} {268} (\bibinfo {year} {1946})}\BibitemShut
  {NoStop}%
\bibitem [{\citenamefont {Radeka}(1974)}]{electronicCooling}%
  \BibitemOpen
  \bibfield  {author} {\bibinfo {author} {\bibfnamefont {V.}~\bibnamefont
  {Radeka}},\ }\href {https://doi.org/10.1109/TNS.1974.4327444} {\bibfield
  {journal} {\bibinfo  {journal} {IEEE Transactions on Nuclear Science}\
  }\textbf {\bibinfo {volume} {21}},\ \bibinfo {pages} {51} (\bibinfo {year}
  {1974})}\BibitemShut {NoStop}%
\bibitem [{\citenamefont {Al~Kenany}\ \emph {et~al.}(2017)\citenamefont
  {Al~Kenany}, \citenamefont {Anil}, \citenamefont {Backes}, \citenamefont
  {Brubaker}, \citenamefont {Cahn}, \citenamefont {Carosi}, \citenamefont
  {Gurevich}, \citenamefont {Kindel}, \citenamefont {Lamoreaux}, \citenamefont
  {Lehnert} \emph {et~al.}}]{haystacDesign}%
  \BibitemOpen
  \bibfield  {author} {\bibinfo {author} {\bibfnamefont {S.}~\bibnamefont
  {Al~Kenany}}, \bibinfo {author} {\bibfnamefont {M.}~\bibnamefont {Anil}},
  \bibinfo {author} {\bibfnamefont {K.}~\bibnamefont {Backes}}, \bibinfo
  {author} {\bibfnamefont {B.}~\bibnamefont {Brubaker}}, \bibinfo {author}
  {\bibfnamefont {S.}~\bibnamefont {Cahn}}, \bibinfo {author} {\bibfnamefont
  {G.}~\bibnamefont {Carosi}}, \bibinfo {author} {\bibfnamefont
  {Y.}~\bibnamefont {Gurevich}}, \bibinfo {author} {\bibfnamefont
  {W.}~\bibnamefont {Kindel}}, \bibinfo {author} {\bibfnamefont
  {S.}~\bibnamefont {Lamoreaux}}, \bibinfo {author} {\bibfnamefont
  {K.}~\bibnamefont {Lehnert}}, \emph {et~al.},\ }\href
  {https://doi.org/10.1016/j.nima.2017.02.012} {\bibfield  {journal} {\bibinfo
  {journal} {Nuclear Instruments and Methods in Physics Research Section A:
  Accelerators, Spectrometers, Detectors and Associated Equipment}\ }\textbf
  {\bibinfo {volume} {854}},\ \bibinfo {pages} {11–24} (\bibinfo {year}
  {2017})}\BibitemShut {NoStop}%
\bibitem [{\citenamefont {Cervantes}\ \emph {et~al.}(2022)\citenamefont
  {Cervantes}, \citenamefont {Braggio}, \citenamefont {Giaccone}, \citenamefont
  {Frolov}, \citenamefont {Grassellino}, \citenamefont {Harnik}, \citenamefont
  {Melnychuk}, \citenamefont {Pilipenko}, \citenamefont {Posen},\ and\
  \citenamefont {Romanenko}}]{SRFC}%
  \BibitemOpen
  \bibfield  {author} {\bibinfo {author} {\bibfnamefont {R.}~\bibnamefont
  {Cervantes}}, \bibinfo {author} {\bibfnamefont {C.}~\bibnamefont {Braggio}},
  \bibinfo {author} {\bibfnamefont {B.}~\bibnamefont {Giaccone}}, \bibinfo
  {author} {\bibfnamefont {D.}~\bibnamefont {Frolov}}, \bibinfo {author}
  {\bibfnamefont {A.}~\bibnamefont {Grassellino}}, \bibinfo {author}
  {\bibfnamefont {R.}~\bibnamefont {Harnik}}, \bibinfo {author} {\bibfnamefont
  {O.}~\bibnamefont {Melnychuk}}, \bibinfo {author} {\bibfnamefont
  {R.}~\bibnamefont {Pilipenko}}, \bibinfo {author} {\bibfnamefont
  {S.}~\bibnamefont {Posen}},\ and\ \bibinfo {author} {\bibfnamefont
  {A.}~\bibnamefont {Romanenko}},\ }\href@noop {} {\bibinfo {title} {Deepest
  sensitivity to wavelike dark photon dark matter with srf cavities}} (\bibinfo
  {year} {2022}),\ \Eprint {https://arxiv.org/abs/2208.03183} {arXiv:2208.03183
  [hep-ex]} \BibitemShut {NoStop}%
\bibitem [{\citenamefont {Brubaker}\ \emph {et~al.}(2017)\citenamefont
  {Brubaker}, \citenamefont {Zhong}, \citenamefont {Lamoreaux}, \citenamefont
  {Lehnert},\ and\ \citenamefont {van Bibber}}]{haystac_2017}%
  \BibitemOpen
  \bibfield  {author} {\bibinfo {author} {\bibfnamefont {B.}~\bibnamefont
  {Brubaker}}, \bibinfo {author} {\bibfnamefont {L.}~\bibnamefont {Zhong}},
  \bibinfo {author} {\bibfnamefont {S.}~\bibnamefont {Lamoreaux}}, \bibinfo
  {author} {\bibfnamefont {K.}~\bibnamefont {Lehnert}},\ and\ \bibinfo {author}
  {\bibfnamefont {K.}~\bibnamefont {van Bibber}},\ }\bibfield  {journal}
  {\bibinfo  {journal} {Physical Review D}\ }\textbf {\bibinfo {volume} {96}},\
  \href {https://doi.org/10.1103/physrevd.96.123008}
  {10.1103/physrevd.96.123008} (\bibinfo {year} {2017})\BibitemShut {NoStop}%
\bibitem [{\citenamefont {Turin}(1960)}]{MatchedFilterIntro}%
  \BibitemOpen
  \bibfield  {author} {\bibinfo {author} {\bibfnamefont {G.}~\bibnamefont
  {Turin}},\ }\href {https://doi.org/10.1109/TIT.1960.1057571} {\bibfield
  {journal} {\bibinfo  {journal} {IRE Transactions on Information Theory}\
  }\textbf {\bibinfo {volume} {6}},\ \bibinfo {pages} {311} (\bibinfo {year}
  {1960})}\BibitemShut {NoStop}%
\bibitem [{\citenamefont {Leffel}\ and\ \citenamefont
  {Daniel}(2021)}]{yfactorRhodeAndSchwartz}%
  \BibitemOpen
  \bibfield  {author} {\bibinfo {author} {\bibfnamefont {M.}~\bibnamefont
  {Leffel}}\ and\ \bibinfo {author} {\bibfnamefont {R.}~\bibnamefont
  {Daniel}},\ }\href {http://www.rohde-schwarz.com/appnote/1MA178} {\bibfield
  {journal} {\bibinfo  {journal} {Rohde \& Schwarz Application Note}\ }
  (\bibinfo {year} {2021})}\BibitemShut {NoStop}%
\bibitem [{\citenamefont {Stutzman}\ and\ \citenamefont
  {Thiele}(1998)}]{Stutzman02}%
  \BibitemOpen
  \bibfield  {author} {\bibinfo {author} {\bibfnamefont {W.~L.}\ \bibnamefont
  {Stutzman}}\ and\ \bibinfo {author} {\bibfnamefont {G.~A.}\ \bibnamefont
  {Thiele}},\ }\href@noop {} {\emph {\bibinfo {title} {Antenna Theory and
  Design}}}\ (\bibinfo  {publisher} {John Wiley Sons},\ \bibinfo {year}
  {1998})\BibitemShut {NoStop}%
\bibitem [{\citenamefont {Kraus}(1950)}]{OriginalKraus}%
  \BibitemOpen
  \bibfield  {author} {\bibinfo {author} {\bibfnamefont {J.}~\bibnamefont
  {Kraus}},\ }\href@noop {} {\emph {\bibinfo {title} {Antennas}}}\ (\bibinfo
  {publisher} {McGraw-Hill},\ \bibinfo {year} {1950})\BibitemShut {NoStop}%
\bibitem [{\citenamefont {Orjubin}\ \emph {et~al.}(2006)\citenamefont
  {Orjubin}, \citenamefont {Richalot}, \citenamefont {Mengue},\ and\
  \citenamefont {Picon}}]{UndermodedRegime}%
  \BibitemOpen
  \bibfield  {author} {\bibinfo {author} {\bibfnamefont {G.}~\bibnamefont
  {Orjubin}}, \bibinfo {author} {\bibfnamefont {E.}~\bibnamefont {Richalot}},
  \bibinfo {author} {\bibfnamefont {S.}~\bibnamefont {Mengue}},\ and\ \bibinfo
  {author} {\bibfnamefont {O.}~\bibnamefont {Picon}},\ }\href
  {https://doi.org/10.1109/TEMC.2006.870705} {\bibfield  {journal} {\bibinfo
  {journal} {IEEE Transactions on Electromagnetic Compatibility}\ }\textbf
  {\bibinfo {volume} {48}},\ \bibinfo {pages} {248} (\bibinfo {year}
  {2006})}\BibitemShut {NoStop}%
\bibitem [{COM(2018)}]{COMSOLRef}%
  \BibitemOpen
  \href@noop {} {\emph {\bibinfo {title} {COMSOL Multiphysics}}},\ \bibinfo
  {organization} {COMSOL, Inc.},\ \bibinfo {address} {One First Street, Suite 4
  Los Altos, CA 94022},\ \bibinfo {edition} {{5.4}}\ ed. (\bibinfo {year}
  {2018}),\ \bibinfo {note} {available at
  \url{https://www.comsol.com/}}\BibitemShut {NoStop}%
\bibitem [{\citenamefont {Zhu}\ \emph {et~al.}(2023)\citenamefont {Zhu},
  \citenamefont {Jewell}, \citenamefont {Laffan}, \citenamefont {Bai},
  \citenamefont {Ghosh}, \citenamefont {Graham}, \citenamefont {Cahn},
  \citenamefont {Maruyama},\ and\ \citenamefont {Lamoreaux}}]{Zhu_2023}%
  \BibitemOpen
  \bibfield  {author} {\bibinfo {author} {\bibfnamefont {Y.}~\bibnamefont
  {Zhu}}, \bibinfo {author} {\bibfnamefont {M.~J.}\ \bibnamefont {Jewell}},
  \bibinfo {author} {\bibfnamefont {C.}~\bibnamefont {Laffan}}, \bibinfo
  {author} {\bibfnamefont {X.}~\bibnamefont {Bai}}, \bibinfo {author}
  {\bibfnamefont {S.}~\bibnamefont {Ghosh}}, \bibinfo {author} {\bibfnamefont
  {E.}~\bibnamefont {Graham}}, \bibinfo {author} {\bibfnamefont {S.~B.}\
  \bibnamefont {Cahn}}, \bibinfo {author} {\bibfnamefont {R.~H.}\ \bibnamefont
  {Maruyama}},\ and\ \bibinfo {author} {\bibfnamefont {S.~K.}\ \bibnamefont
  {Lamoreaux}},\ }\bibfield  {journal} {\bibinfo  {journal} {Review of
  Scientific Instruments}\ }\textbf {\bibinfo {volume} {94}},\ \href
  {https://doi.org/10.1063/5.0137870} {10.1063/5.0137870} (\bibinfo {year}
  {2023})\BibitemShut {NoStop}%
\bibitem [{\citenamefont {Caputo}\ \emph {et~al.}(2021)\citenamefont {Caputo},
  \citenamefont {Millar}, \citenamefont {O'Hare},\ and\ \citenamefont
  {Vitagliano}}]{DarkPhotonHandbook}%
  \BibitemOpen
  \bibfield  {author} {\bibinfo {author} {\bibfnamefont {A.}~\bibnamefont
  {Caputo}}, \bibinfo {author} {\bibfnamefont {A.~J.}\ \bibnamefont {Millar}},
  \bibinfo {author} {\bibfnamefont {C.~A.~J.}\ \bibnamefont {O'Hare}},\ and\
  \bibinfo {author} {\bibfnamefont {E.}~\bibnamefont {Vitagliano}},\ }\href
  {https://doi.org/10.1103/PhysRevD.104.095029} {\bibfield  {journal} {\bibinfo
   {journal} {Phys. Rev. D}\ }\textbf {\bibinfo {volume} {104}},\ \bibinfo
  {pages} {095029} (\bibinfo {year} {2021})}\BibitemShut {NoStop}%
\end{thebibliography}%

\end{document}